\documentclass[12pt]{iopart}
\usepackage{iopams}
\usepackage{graphicx}
\usepackage{color}
\usepackage{hyperref}
\usepackage{tikz}
\usepackage{url}
\usepackage[T1]{fontenc}

\newcommand{\Rs}{R_{\mathrm{S}}}
\newcommand{\ka}{\kappa}

\newcommand{\ten}{R/\Rs}
\newcommand{\Rb}{R_{\mathrm{B}}}
\newcommand{\ee}{\mathrm{e}}

\definecolor{CPcolor}{rgb}{0.2, 0.13, 0.9}

\definecolor{PBcolor}{rgb}{0.9, 0.5, 0.0}

\eqnobysec
\begin{document}

\title[Slowly rotating ultra-compact Schwarzschild star in the gravastar limit]{Slowly rotating ultra-compact Schwarzschild star in the gravastar limit}

\author{Philip Beltracchi and Camilo Posada$^{1,2}$}
\address{$^1$Research Centre for Theoretical Physics and Astrophysics, Institute of Physics, Silesian University in Opava, Bezru\v{c}ovo n\'{a}m. 13, CZ-746 01 Opava, Czech Republic}
\address{$^2$Programa de Matem\'atica, Fundaci\'on Universitaria Konrad Lorenz, 110231 Bogot\'a, Colombia}

\ead{phipbel@aol.com and camilo.posada@physics.slu.cz}

\begin{abstract}

We reconsider the problem of a slowly rotating homogeneous star, or Schwarzschild star, when its compactness goes beyond the Buchdahl bound and approaches the gravastar limit $R\to 2M$. We compute surface and integral properties of such configuration by integrating the Hartle-Thorne structure equations for slowly rotating relativistic masses, at second order in angular velocity. In the gravastar limit, we show that the metric of a slowly rotating Schwarzschild star agrees with the Kerr metric, thus, within this approximation, it is not possible to tell a gravastar from a Kerr black hole by any observations from the spacetime exterior to the horizon.

\end{abstract}



\section{Introduction}

It has been known for a long time, that a star whose mass exceeds some critical value, and which is contained in a sufficiently small region, will develop trapped surfaces. Under certain conditions, this will subsequently lead to gravitational collapse to a singular state called a ‘black hole' (BH) \cite{Hawking:1973uf}. These results are known as the singularity theorems \cite{Penrose:1964wq,Hawking:1970zqf}. The BH has been the source of plethora of investigations in the last 60 years (see e.g~\cite{Wald:1999vt,Chrusciel:2012jk,Barack:2018yly}, and references therein), reaching recently a peak with the LIGO/Virgo detection of gravitational waves signals of the merger of two BHs~\cite{LIGOScientific:2016aoc, LIGOScientific:2016sjg}, and the outstanding images of M87~\cite{EventHorizonTelescope:2019dse} and Sagittarius $\mathrm{A}^{*}$~\cite{EventHorizonTelescope:2022wkp}, provided by the Event Horizon Telescope collaboration.

One of the key features of mathematical BHs is the so-called event horizon, which is a marginally trapped surface where no information whatsoever can escape \cite{Hawking:1973uf}. This event horizon, has been the root of severe issues in BH physics, like the ‘BH information paradox'~\cite{Hawking:1976ra}, which despite many proposals (see e.g.~\cite{Chen:2014jwq} and references therein), remains as one of the biggest challenges of the BH paradigm. Although the exterior geometry of BHs is free of pathologies, and it has been thoroughly studied~\cite{Chandrasekhar:1983kt}, the interior is not; for instance, standard BHs hold curvature singularities in their interior which breaks the classical concepts of space and time as well as the known laws of physics \cite{Hawking:1976ra}. Curvature singularities are a feature of the interior for the standard Schwarzschild, Reissner-Nordstr\"om, Kerr, and Kerr-Newman models of BHs~\cite{Chandrasekhar:1983kt}. Additionally, Kerr BHs predict the existence of acausal closed timelike geodesics in their interior~\cite{Hawking:1973uf}.

Certain models of non-singular BHs, or BH mimickers, have been introduced in the literature, as a way out to the BH paradoxes~\cite{Cardoso:2019rvt}. These objects avoid the appearance of singularities by violating one or more of the assumptions of the singularity theorems. One of the most popular models is the gravastar proposed by Mazur and Mottola~\cite{Mazur:2001fv,Mazur:2004fk}. The gravastar is characterized by an interior de Sitter region with a positive (and constant) energy density $\epsilon$, and a negative pressure $p=-\epsilon$, which is matched to the exterior Schwarzschild spacetime by a timelike thin-shell boundary of thickness $l\sim \left(L_{\mathrm{Pl}}r_{M}\right)^{1/2}$, where $L_{\mathrm{Pl}}$ is the Planck length, and $r_{M}\sim r_{\mathrm{H}}$ is the boundary of the mutual de Sitter and Schwarzschild horizons. This collapsed cold dark object, has no event horizon, a regular interior, and a globally defined Killing time. Moreover, in contrast with classical BHs, the gravastar does not suffer from huge entropies and information paradox.

Various generalizations of the gravastar model have been under scrutiny during the last 20 years~\cite{Visser:2003ge,Carter:2005pi,Cattoen:2005he,DeBenedictis:2005vp,Lobo:2006xt,Chirenti:2007mk,Pani:2009ss,Pani:2010em,Cardoso:2016oxy,Chirenti:2008pf,Cardoso:2007az,Pani:2015tga,Uchikata:2016qku,Uchikata:2015yma,Chirenti:2016hzd,Rutkowski:2020fvm,Das:2017rhi}.
Gravitational wave signatures and non-radial oscillations of thin-shell gravastars were studied by~\cite{Chirenti:2007mk,Pani:2009ss,Pani:2010em,Cardoso:2016oxy}. Some models of rotating gravastars have been proposed by a number of authors~\cite{Chirenti:2008pf,Cardoso:2007az,Pani:2015tga,Uchikata:2016qku,Uchikata:2015yma,Chirenti:2016hzd,Rutkowski:2020fvm}. In this context, the ergoregion instability of rotating gravastars, which affects compact objects rotating rapidly, was investigated by~\cite{Chirenti:2008pf,Cardoso:2007az}. For instance, \cite{Chirenti:2008pf} found that some stable gravastars can be constructed with $J/M^2\sim 1$, with $J$ being the angular momentum and $M$ is the total mass; thus not all of the astrophysical objects rotating with $J/M^2 <\sim 1$ are to be considered BHs.

The gravastar as pictured originally in \cite{Mazur:2001fv,Mazur:2004fk}, has been reexamined by \cite{Mazur:2015kia}, after the realization that an infinitesimally thin-shell gravastar emerges naturally from the Schwarzschild's interior solution with constant density~\cite{Schwarzschild:1916inc}, or ‘Schwarzschild star'. It is well-known that the central pressure of the Schwarzschild star diverges when its radius reaches the Buchdahl bound $\Rb\equiv(9/4)M$. Thus, the constant density solution saturates the Buchdahl theorem~\cite{Buchdahl:1959zz}, which establishes that the radius of a star must satisfy the inequality $R\geq (9/4)M$, in order to be in hydrostatic equilibrium, regardless of the equation of state, if the pressure is positive and isotropic.

By contracting adiabatically the Schwarzschild star, beyond the Buchdahl bound, \cite{Mazur:2015kia} showed that the central pressure divergence moves outward from the origin to a surface $R_{0}=3R\sqrt{1-(8/9)(R/\Rs)}$ and a negative pressure region opens up behind it. In the limit when the radius of the star approaches the Schwarzschild radius $\Rs\equiv 2M$, the Schwarzschild star produces a $p=-\epsilon$ interior, which is matched to the exterior Schwarzschild spacetime by an infinitesimally thin (null) boundary layer of anisotropic stresses localized at $\Rs$. Thus, in this limit, the Schwarzschild star becomes essentially the gravastar. 

The Schwarzschild star has been shown to be stable against radial oscillations~\cite{Posada:2018agb}. Axial non-spherical perturbations of ultracompact Schwarzschild stars were considered by~\cite{Konoplya:2019nzp}. These authors showed that the observable quasi-normal modes can be as close as one wishes to the Schwarzschild limit, making it a very good BH mimicker. Moreover, the decaying time-domain profile shows that the ultracompact Schwarzschild star is stable against axial oscillations (although no echoes were found). An exact solution for dynamical collapse of a modified anisotropic Schwarzschild star was found by \cite{Beltracchi:2019tsu}. Chirenti et.~al.~\cite{Chirenti:2020bas} determined the tidal Love number $k_2$ of ‘sub-Buchdahl' Schwarzschild stars, and they found that it is strongly dependent on the compactness $M/R$; moreover, in the Schwarzschild limit, $k_{2}\to 0$, thus a gravastar cannot be distinguished from a BH by its tidal deformability.

A first extension of the ultracompact Schwarzschild star to slow rotation was proposed in~\cite{Posada:2016xxx} by employing the Hartle-Thorne (HT)~\cite{Hartle:1967he,Hartle:1968si} perturbative approach for computing the properties of slowly rotating relativistic stars to second order in the angular velocity, together with the methods introduced by Chandrasekhar and Miller~\cite{Chandra:1974} for slowly rotating Schwarzschild stars, below the Buchdahl limit. However, as it was pointed out by~\cite{Beltracchi:2021lez}, the results reported by~\cite{Posada:2016xxx}, beyond the Buchdahl limit, were marred after an incorrect assumption in a coordinate transformation. While it is possible to solve Hartle's equations for dark energy analytically~\cite{Pani:2015tga, Uchikata:2016qku, Beltracchi:2021lez,Uchikata:2014kwm}, the $R\rightarrow 2M$ limit of  numerically computed slowly rotating Schwarzschild stars can provide a valuable point of comparison to these analytic results, especially to those of \cite{Beltracchi:2021lez} which used a null boundary and the same interior time scaling of the Schwarzschild derived gravastar from \cite{Mazur:2015kia}.

In this paper we reconsider the problem studied in \cite{Posada:2016xxx} of a slowly rotating Schwarzschild star when its compactness goes beyond the Buchdahl bound and approaches the gravastar limit $R\to \Rs$. We follow the formulation of~\cite{Chandra:1974}, where the HT structure equations were integrated for a Schwarzschild star up to the Buchdahl limit. Our approach will complete, and improve, the one presented by~\cite{Posada:2016xxx} in various directions. First, we apply boundary conditions of the HT equations at $R_{0}$, rather than the origin \cite{Beltracchi:2021}, which allowed us to compute surface and integral properties (i.e., physical properties of the object like moment of inertia, change in mass, etc., evaluated at the surface) in the regime $\Rs<R<(9/8)\Rs$. Second, we discuss the inward extension of the solution to the perturbation equation, inside the infinite surface pressure surface $R_0$. We found that in the gravastar limit, the perturbation functions and quantities like the moment of inertia, and mass quadrupole moment, approach the corresponding Kerr BH values. Thus, our results indicate that the exterior spacetime of a slowly rotating gravastar cannot be differentiated to that of a slowly rotating Kerr BH. 

The paper is organized as follows: in section~\ref{Sect:2} we review the Schwarzschild interior solution for incompressible fluids, or Schwarzschild star, and its approach to the gravastar limit. In section~\ref{Sect:3} we summarize the general Hartle-Thorne framework for slowly rotating relativistic stars. In section~\ref{Sect:4} we apply the HT approximation to the Schwarzschild stars. In section~\ref{Sect:5} we present our methods and results of surface and integral properties, where in section~\ref{Sect:5.1} we discuss our results for frame dragging and moment of inertia. Results related to the monopole perturbations, such as the properties of $m_0$, $h_0$, $p_0^*$, $\xi_0$, and the change in mass $\delta M$  are presented in section~\ref{Sect:5.2}. The properties of the quadrupole perturbations $m_2$, $h_2$, $k_2$, and $p_2^*$, together with the external parameters $\xi_2$, mass quadrupole moment $Q$, and ellipticity $\epsilon$, as well as the $I-Q$ relation, are presented in section~\ref{Sect:5.3}. Section~\ref{Sect:6} contains our final conclusions. We use geometrized units $(G=c=1)$, unless stated otherwise, and we adopt the metric signature $(-,+,+,+)$.

There are three appendices. The first contains a series solution to the ‘dragging' equation, where we argue on the choice of the branch $\varpi_1$. In \ref{appendix2} we provide a consistency check between the $(r,R)$ and $(\ka,x)$ coordinates. Finally, in \ref{appendix3}, we determine the surface gravity and angular momentum parameters of the slowly rotating Schwarzschild stars.

\section{Spherically symmetric Schwarzschild star in the gravastar limit}\label{Sect:2}
\subsection{Non-rotating configuration in hydrostatic equilibrium}\label{Sect:2a}

The nonrotating configuration is described by a metric in the standard spherically symmetric form 
 \begin{eqnarray}\label{statmetric}
ds^2 = -\ee^{2\nu(r)}dt^2+\ee^{2\lambda(r)}dr^2 + r^2 (d\theta^2 + \sin^2\theta d\phi^2),
\end{eqnarray}
where $\nu$ and $\lambda$ are only functions of $r$. We assume a perfect fluid with an isobaric equation of state, i.e., $\epsilon=\epsilon(p)$, which connects the pressure $p$ with the mass-energy density $\epsilon$. For a given value of the central energy density $\epsilon_{\mathrm{c}}$, the nonrotating configuration is determined by solving the Tolman-Oppenheimer-Volkoff (TOV) equation of hydrostatic equilibrium for the pressure $p(r)$, and the mass enclosed by a given radius $m(r)$
\begin{eqnarray}\label{tov}
\frac{dp}{dr}=-(\epsilon+p)\frac{m(r)+4\pi pr^3}{r[r-2m(r)]},
\end{eqnarray}
\begin{eqnarray}\label{misner}
\frac{dm(r)}{dr}=4\pi\epsilon(r)r^2.
\end{eqnarray}
The mass function $m(r)$ is connected with the metric component $g_{rr}$ by the relation
\begin{eqnarray}\label{grr0}
 \ee^{-2\lambda(r)}\equiv 1-\frac{2m(r)}{r},
\end{eqnarray}
\noindent and the time-component metric $g_{tt}$ can be obtained as
\begin{eqnarray}\label{gtt0}
\frac{d\nu}{dr}=-\frac{1}{(\epsilon+p)}\left(\frac{dp}{dr}\right).
\end{eqnarray}
\noindent In the exterior vacuum space-time $\epsilon=p=0$, thus the geometry is described by the Schwarzschild metric
\begin{eqnarray}\label{ext_schw}
 \ee^{2\nu(r)}=\ee^{-2\lambda(r)}=1-\frac{2M}{r},\quad r>R, 
\end{eqnarray}
\noindent where $R$ is the radius and $m(R)=M$ is the total gravitational mass of the configuration. The interior and exterior spacetimes are matched across the boundary $\Sigma=R$, such that 
\begin{eqnarray}
[\lambda]=0,\quad [\nu]=0,\quad [\nu']=0,
\end{eqnarray}
\noindent where $[f]$ indicates the difference between the value of $f$ in the vacuum exterior and its value in the interior, evaluated at $\Sigma$, i.e., $[f]=f^{+}\vert_{\Sigma}-f^{-}\vert_{\Sigma}$, and the prime denotes derivative with respect to the radial coordinate $r$ \footnote{We use the opposite convention from \cite{Reina:2015jia}, who uses $+ (-)$ to denote quantities in the interior (exterior).}.

\subsection{Schwarzschild's interior solution in the gravastar limit}\label{Sect:2b}

If we model the star as an incompressible and isotropic fluid with
\begin{eqnarray}\label{rho}
\epsilon = \frac{3M}{4\pi R^3}=\frac{3H^2}{8\pi}=\mathrm{constant},
\end{eqnarray}
\noindent where $H^2=\Rs/R^3$, by solving equations \eref{tov}-\eref{gtt0} we obtain the well-known Schwarzschild's interior solution with constant density \cite{Schwarzschild:1916inc} (a.k.a. Schwarzschild star) 
\begin{eqnarray}\label{gttSstar}
\ee^{2\nu(r)} = \frac{1}{4}\left[3\sqrt{1-H^2R^2}-\sqrt{1-H^2r^2}\right]^2,
\end{eqnarray} 
\begin{eqnarray}\label{grrSstar}
\ee^{-2\lambda(r)}=1-H^2r^2,
\end{eqnarray}
\begin{eqnarray}\label{pSstar}
p(r) = \epsilon\left[\frac{\sqrt{1-H^2r^2} - \sqrt{1-H^2R^2}}{3\sqrt{1-H^2R^2}-\sqrt{1-H^2r^2}}\right].
\end{eqnarray} 
\noindent At the boundary of the configuration $r=R$, this solution must match with the exterior solution \eref{ext_schw}, and the pressure must vanish $p(R)=0$. The continuity of $g_{tt}$ guarantees that an observer crossing the boundary will not notice any discontinuity of time measurements. 

From \eref{pSstar} we note that the central pressure blows up when the stellar radius reaches the Buchdahl limit $\Rb\equiv(9/8)\Rs$. As it was shown by Buchdahl~\cite{Buchdahl:1959zz}, under certain assumptions, this limit is actually independent of the equation of state describing the fluid. Thus, in order to be in hydrostatic equilibrium the radius of a star must satisfy the inequality $R\geq (9/8)\Rs$. In connection with this, we observe that \eref{pSstar} is regular except at some radius $R_{0}$ where the denominator
\begin{eqnarray}\label{denom}
D(r) \equiv 3\sqrt{1-H^2R^2}-\sqrt{1-H^2r^2},
\end{eqnarray} 
\noindent vanishes in the range $0<r<R$. Remarkably, it can be seen from \eref{gttSstar} and \eref{pSstar} that the pressure goes to infinity at the same point where $\ee^{2\nu}=0$. This singular radius can be found directly from \eref{denom} as
\begin{eqnarray}\label{r0}
R_{0} = 3R\sqrt{1-\frac{8}{9}\frac{R}{\Rs}},
\end{eqnarray} 
\noindent which is imaginary for $R$ above the Buchdahl limit. As $R\to (9/8)\Rs$, from above, \eref{r0} shows that $R_{0}$ approaches the real axis at $R_{0}=0$ and a divergence of the pressure appears as $\ee^{2\nu}\to 0$. Note that when $R=\Rb$, $R_0=0$, thus the pole appears first at the origin. Mazur and Mottola \cite{Mazur:2015kia} studied further the region beyond the Buchdahl limit, i.e. $\Rs<R<(9/8)\Rs$, and they found that $R_{0}$ moves outwards from the origin to finite values $0<R_{0}<R$ (see figure~1 in \cite{Posada:2016xxx}). Thus, there emerges naturally a new region  $0\leq r < R_{0}$, where $D(r)<0$, and in consequence $p(r)<0$, while $\ee^{2\nu}>0$. As the adiabatic and quasi-stationary contraction continues from above, i.e., $R\to \Rs^{+}$, $R_{0}\to \Rs^{-}$ from below thus the region $0<r<R_{0}$ with negative pressure keeps growing while the region with positive pressure $R_{0}<r<R$ squeezes (see figure~3 in \cite{Posada:2016xxx}). In the limit when $R=R_{0}=\Rs$, the Schwarzschild's interior solution \eref{gttSstar}, \eref{grrSstar} and \eref{pSstar} becomes
\begin{eqnarray}\label{interior}
\ee^{2\nu(r)} = \frac{1}{4}\left(1 - \frac{r^2}{\Rs^2}\right),\quad\ee^{2\lambda(r)} = \left(1-\frac{r^2}{\Rs^2}\right)^{-1},
\end{eqnarray}
\begin{eqnarray}
p = -\epsilon,
\end{eqnarray}
\noindent which is essentially the gravastar proposed in \cite{Mazur:2015kia}, parametrized by the total mass $M$. The exterior region $r>\Rs$ remains the vacuum spherically symmetric Schwarzschild geometry \eref{ext_schw}, with an infinitesimal thin shell discontinuity at the surface $\Rs$ where there is a jump in pressure and the zeroes $\ee^{2\nu}=\ee^{2\lambda}=0$ of the interior de Sitter and exterior Schwarzschild spacetimes match.

The divergence in pressure at $R_{0}$ can be integrated through the Komar formula, if one assumes that $p_{\perp}\neq p$, thus relaxing the isotropic fluid condition. Beyond the Buchdahl bound, $R/\Rs<9/8$, the relation for pressure is \cite{Mazur:2015kia}
\begin{eqnarray}\label{delta}  
8\pi\sqrt{\frac{f}{h}}r^2(p_{\perp}-p)=\frac{8\pi\epsilon}{3} R_{0}^3\delta(r-R_{0}),
\end{eqnarray}
\noindent indicating the presence of a $\delta$-distributional anisotropic stress tensor at $r=R_{0}$, which produces a surface tension 
\begin{eqnarray}\label{tension}  
\tau_{s}=\frac{[\kappa]}{8\pi}=\frac{1}{16\pi M},
\end{eqnarray}
\noindent that is proportional to the difference in surface gravities
\begin{eqnarray}\label{deltak}  
[\kappa]\equiv\kappa_{+}-\kappa_{-}=\frac{\Rs R_0}{R^3}.
\end{eqnarray}
\noindent In contrast to a BH, \eref{tension} corresponds to a physical surface tension (localized in an infinitesimal thin shell at $r=\Rs$) provided by a positive transverse pressure as determined by the Komar formula. The boundary $r=\Rs$ is a null hypersurface, but it is not an event horizon, thus there is no ‘information paradox'. Moreover, there are no trapped surfaces in the region $r<\Rs$, and it has a regular interior free of singularities. Furthermore, the gravastar is a cold condensate state with low entropy in contrast to BHs which possesses huge entropies as predicted by the Bekenstein-Hawking formula, far in excess of the stellar progenitors~\cite{Mazur:2001fv}. 

\section{Hartle-Thorne formalism for rotational metric deformations}
\label{Sect:3}

 Once the equilibrium configuration is set into slow rotation, the interior distribution and the spacetime geometry change. The appropriate line element for this situation is the so-called Hartle-Thorne metric \cite{Hartle:1967he, Hartle:1968si}
\begin{eqnarray}\label{axialmetric}
ds^2 =&~-\ee^{2\nu}\left[1+2h_{0}(r)+2h_{2}(r)P_{2}(\cos\theta)\right]dt^2\nonumber \\
&+\ee^{2\lambda}\left\{1+\frac{\ee^{2\lambda}}{r}\left[2m_{0}(r) + 2m_{2}(r)P_{2}(\cos\theta)\right]\right\}dr^2 \nonumber\\
&+r^2\left[1+2k_{2}(r)P_{2}(\cos\theta)\right]\left[d\theta^2 +\sin^2\theta (d\phi - \omega dt)^2\right],
\end{eqnarray}
\noindent where $P_{2}(\cos\theta)$ is the Legendre polynomial of order 2; $(h_{0},h_{2},m_{0},m_{2},k_{2})$ are quantities of order $\Omega^2$; and $\omega(r)$ (of order $\Omega$) is the ‘dragging' function corresponding to the angular velocity of the local inertial frame, relative to a distant observer. The angular velocity of the star, as measured by distant observers, is indicated by $\Omega$. In the nonrotating case, the metric \eref{axialmetric} reduces to \eref{statmetric}.

\subsection{Moment of inertia and dragging of inertial frames}

We assume the fluid, inside the configuration, rotating uniformly with four-velocity $u^{\mu}$ given by 
\begin{eqnarray}
u^{t} &= (-g_{tt} - 2\Omega g_{t\phi} - g_{\phi\phi}\Omega^2)^{-1/2} \nonumber \\
&= e^{-\nu}\left[1+\frac{1}{2}r^2\sin^2\theta(\Omega-\omega)^2e^{-2\nu}-h_{0}-h_{2}P_{2} \right], \nonumber\\
u^{\phi} &= \Omega u^{t},\quad u^{r}=u^{\theta}=0 \nonumber.
\end{eqnarray}
\noindent It is conventional to introduce the quantity 
\begin{eqnarray}
\varpi\equiv \Omega - \omega,
\end{eqnarray}
\noindent that corresponds to the angular velocity of the fluid relative to the local inertial frame. This quantity determines the centrifugal force and, at first order in $\Omega$, satisfies the following equation~\cite{Hartle:1967he,Hartle:1968si}
\begin{eqnarray}\label{omegain}
\frac{d}{dr}\left[r^4j(r)\frac{d\varpi}{dr}\right]+4r^3\frac{dj}{dr}\varpi=0,
\end{eqnarray}
\noindent where
\begin{eqnarray}\label{j}
j(r)\equiv e^{-(\lambda+\nu)}.
\end{eqnarray}
\noindent In the region $r>R$, $\epsilon=p=0$ and the spacetime geometry is described by \eref{ext_schw}; thus, $j(r)=1$ and \eref{omegain} can be easily integrated to give
\begin{eqnarray}\label{omegaout}
\varpi(r)^{+}=\Omega - \frac{2J}{r^3},
\end{eqnarray}
\noindent where $J$ is an integration constant associated with the angular momentum of the star \cite{Hartle:1967he}. For nonsingular well behaved solutions, it can be proved, using series arguments, that the appropriate boundary conditions for \eref{omegain} are, $\varpi(0)=\varpi_{\mathrm{c}}=\mathrm{const}.$, and $(d\varpi/dr)\vert_{r=0}=0$, after which it is integrated outward.~Regularity demands that, at the surface, the interior and exterior solutions, together with their derivatives, must match, i.e., $[\varpi]=[\varpi']=0$. Thus, one integrates numerically \eref{omegain} and obtains the surface value $\varpi(R)$, then, and only then, one can determine the angular momentum $J$ and the angular velocity $\Omega$ as
\begin{eqnarray}\label{wsurf}
J=\frac{1}{6}R^4\left(\frac{d\varpi}{dr}\right)_{r=R},\quad \Omega = \varpi(R) + \frac{2J}{R^3}.
\end{eqnarray}
Once the angular momentum and angular velocity are determined, the relativistic moment of inertia can be obtained by the relation $I = J/\Omega$.

\subsection{Spherical deformations of the star: $l=0$ sector}

In the interior of the star, the pressure can be expanded as
\begin{eqnarray}\label{p_exp}
p_{\mathrm{tot}}(r) = p + (\epsilon+p)\left[p_{0}^{*} + p_{2}^{*} P_{2}(\cos\theta)\right],
\end{eqnarray}
\noindent where $p_{\mathrm{tot}}$ is the pressure in the rotating system, $\epsilon$ and $p$ are the density and pressure in the background system, and $p_{0}^{*}$ and $p_{2}^{*}$ are functions of $r$ proportional to $\Omega^2$, associated to the perturbations in the pressure. 

The equations for the $l=0$ sector, for the mass perturbation factor $m_{0}$, and pressure perturbation factor $p_{0}^{*}$ read \cite{Hartle:1968si,Chandra:1974}
\begin{eqnarray}\label{m0int}
\frac{dm_{0}}{dr} =&4\pi r^2(\epsilon+p)\left(\frac{d\epsilon}{dp}\right)p_{0}^{*} + \frac{1}{12}r^4j^2\left(\frac{d\varpi}{dr}\right)^2 -\frac{1}{3}r^3\varpi^2\frac{dj^2}{dr},
\end{eqnarray}
\begin{eqnarray}\label{dp0int}
\frac{dp_{0}^{*}}{dr} =&-\frac{1+8\pi p r^2}{(r-2m)^2}m_{0}-\frac{4\pi(\epsilon+p)r^2}{r-2m}p_{0}^{*} \nonumber +\frac{1}{12}\frac{r^4 j^2}{(r-2m)}\left(\frac{d\varpi}{dr}\right)^2\nonumber\\
&+\frac{1}{3}\frac{d}{dr}\left(\frac{r^3 j^2 \varpi^2}{r-2m} \right).
\end{eqnarray}
These equations are integrated outward with the boundary conditions $m_{0}=p^{*}_{0}=0$ at the origin, with $m_0(0)=0$ indicating the lack of a point mass term to the perturbations, and $p_0^*(0)=0$ being conventional in the formalism~\cite{Hartle:1967he,Hartle:1968si}. The function $h_{0}$ in the interior of the configuration can be determined from the equation of hydrostatic equilibrium as follows \cite{Hartle:1968si,Chandra:1974}
\begin{eqnarray}\label{h0in}
h_{0}^{-}=-p_{0}^{*}+\frac{1}{3}r^2 \ee^{-2\nu}\varpi^2 + h_\mathrm{c},
\end{eqnarray}
\noindent where $h_\mathrm{c}$ is a constant of integration. In the exterior of the star, where $\epsilon=p=0$, \eref{m0int} and \eref{dp0int} are integrated explicitly giving
\begin{eqnarray}\label{m0out}
m_{0}^{+}(r) = \delta M - \frac{J^2}{r^3},
\end{eqnarray}
\begin{eqnarray}\label{h0out}
h_{0}^{+}(r) = -\frac{\delta M}{r-2M} + \frac{J^2}{r^3(r-2M)},
\end{eqnarray}\\
\noindent where $\delta M$ is a constant of integration associated with the change in mass induced by the rotation. The constants $h_\mathrm{c}$ and $\delta M$ can be found by the continuity of the solutions at the surface $\Sigma$, i.e., $[h_{0}]=[m_{0}]=0$. 

As it was pointed out by \cite{Reina:2014fga}, expression \eref{m0out} for $\delta M$ is not the most general one. By reconsidering the matching problem in Hartle's framework, by employing the perturbed matching theory at second order developed by \cite{Mars:2005ca}, \cite{Reina:2014fga} showed that the perturbative functions are continuous across the surface $\Sigma$, except for the function $m_{0}(r)$ in the $l=0$ sector. The discontinuity in $m_{0}(r)$ turns out to be proportional to the energy density at $\Sigma$, thus contributing an additional term to \eref{m0out} in the form
\begin{eqnarray}\label{m0RV}
\delta M_{\mathrm{mod}} = m_{0}(R) + \frac{J^2}{R^3} + 8\pi R^3\left(\frac{R}{2M}-1\right)\epsilon(R)p_{0}^{*},
\end{eqnarray}
\noindent such that the total mass of the star is $M(R)+\delta M_{\mathrm{mod}}$. Note that for configurations with a vanishing surface energy density, as is the case for most of the equations of state for realistic NSs, this additional term vanishes; thus the original expression \eref{m0out} remains valid. However, for configurations with a discontinuity in $\epsilon$ at the surface, e.g. constant density stars or strange stars, this additional term provides a significant contribution to the change in mass \cite{Reina:2015jia}.

\subsection{Quadrupole deformations of the star: $l=2$ sector}

The quadrupole deformations of the star are determined by the $l=2$ perturbation equations given by \cite{Hartle:1968si,Chandra:1974} 
\begin{eqnarray}\label{v2in}
\frac{dv_{2}}{dr}= -2\left(\frac{d\nu}{dr}\right)h_{2} +\left(\frac{1}{r}+\frac{d\nu}{dr}\right)\left[\frac{1}{6}r^4j^2\left(\frac{d\varpi}{dr}\right)^2-\frac{1}{3}r^3\varpi^2\frac{dj^2}{dr}\right],
\end{eqnarray}
\begin{eqnarray}\label{h2in}
\frac{dh_{2}}{dr} =&~-\frac{2v_{2}}{r\left[r-2m(r)\right]}\left(\frac{d\nu}{dr}\right)^{-1} + \nonumber\\
& \left\{-2\frac{d\nu}{dr}+\frac{r}{2\left[r-2m(r)\right]}\left(\frac{d\nu}{dr}\right)^{-1}\left[8\pi(\epsilon+p)-\frac{4m(r)}{r^3}\right]\right\}h_{2} \nonumber\\
& +\frac{1}{6}\left[r\left(\frac{d\nu}{dr}\right)-\frac{1}{2\left[r-2m(r)\right]}\left(\frac{d\nu}{dr}\right)^{-1}\right]r^3j^2\left(\frac{d\varpi}{dr}\right)^2 \nonumber\\ 
& -\frac{1}{3}\left[r\left(\frac{d\nu}{dr}\right)+\frac{1}{2\left[r-2m(r)\right]}\left(\frac{d\nu}{dr}\right)^{-1}\right]r^2\frac{dj^2}{dr}\varpi^2,
\end{eqnarray}
\noindent where 
\begin{eqnarray}
k_{2}=v_{2}-h_{2}.
\end{eqnarray}
In the exterior where $\epsilon=p=0$, equations \eref{v2in} and \eref{h2in} can be integrated analytically to give 
\begin{eqnarray}\label{h2out}
h_{2}^{+}(r)=\frac{J^2}{Mr^3}\left(1 + \frac{M}{r}\right) + KQ_{2}^{\;2}(\zeta),
\end{eqnarray}
\begin{eqnarray}\label{v2out}
v_{2}^{+}(r)=-\frac{J^2}{r^4} + K\frac{2M}{\left[r(r-2M)\right]^{1/2}}Q_{2}^{\;1}(\zeta),
\end{eqnarray}
\noindent where $K$ is a constant of integration and $Q_{n}^{\;m}$ are the associated Legendre functions of the second kind with argument $\zeta\equiv (r/M) -1$. The condition of regularity of the solutions at the surface $\Sigma$ requires $[h_2]=[v_2]=0$. The constant $K$ appearing in \eref{h2out} and \eref{v2out}, is connected with the quadrupole mass moment of the star, as measured at infinity, via the relation \cite{Hartle:1968si}
\begin{eqnarray}\label{Q}
Q = \frac{J^2}{M} + \frac{8}{5}KM^3.
\end{eqnarray}
To find the constant $K$, the functions $h_2$ and $v_2$ must be computed for the interior of the star, and then matched at the surface $\Sigma$ with the exterior solutions \eref{h2out} and \eref{v2out}.

The perturbations for mass and pressure $m_{2}$ and $p_{2}^{*}$, which determine the rotational deformations of the star, can be determined using the following relations \cite{Hartle:1968si,Chandra:1974}
\begin{eqnarray}\label{m2in}
m_{2} =&~\left(r-2m\right)\left[-h_{2}(r)-\frac{1}{3}r^3\left(\frac{dj^2}{dr}\right)\varpi^2 + \frac{1}{6}r^4j^2\left(\frac{d\varpi}{dr}\right)^2\right],
\end{eqnarray}
\begin{eqnarray}\label{p2in}
p_{2}^{*} = -h_{2(r)} - \frac{1}{3}r^2 \ee^{-2\nu} \varpi^2.
\end{eqnarray}
\noindent The rotational perturbations will displace the nonrotating surface to the radius \cite{Hartle:1967he, Hartle:1968si, Chandra:1974}
\begin{eqnarray}
r^{*}(p) = r + \xi_{0}(r) + \xi_2(r)P_{2}(\cos\theta), 
\end{eqnarray}
\noindent where $r$ is the radius of the nonrotating configuration  with pressure $p$; $\xi_0$ and $\xi_2$ are perturbation functions which satisfy the following relations \cite{Hartle:1968si,Chandra:1974}
\begin{eqnarray}\label{xis}
p_{0}^{*}=-\left[\frac{dp/dr}{(\epsilon+p)}\right]_{0}\xi_{0}(r),\quad p_{2}^{*}=-\left[\frac{dp/dr}{(\epsilon+p)}\right]_{0}\xi_{2}(r),
\end{eqnarray}
\noindent where the terms in square brackets are evaluated in the nonrotating configuration. Using \eref{gtt0} we can rewrite \eref{xis} in the following form
\begin{eqnarray} \label{xi02r}
\xi_{0}(r) = \left(\frac{d\nu}{dr}\right)^{-1}p_{0}^{*},\quad
\xi_{2}(r)=\left(\frac{d\nu}{dr}\right)^{-1}p_{2}^{*}.
\end{eqnarray}
The ellipticity of the isobaric surfaces is determined by \cite{Chandra:1974}
\begin{eqnarray}\label{ellipr}
\varepsilon(r)=-\frac{3}{2r}\left[\xi_{2}(r) + r\left(v_2-h_2\right)\right].
\end{eqnarray}
In the following sections, we will apply the HT approximation to the Schwarzschild star, and study its approach to the gravastar limit.

\section{Hartle-Thorne framework for slowly rotating Schwarzschild stars}\label{Sect:4}

Here we summarize the HT structure equations for slowly rotating Schwarzschild stars, as described by the interior Schwarzschild's solution with uniform density, which we discussed in section~\ref{Sect:2b}. Following the conventions of \cite{Chandra:1974}, equations \eref{gttSstar}-\eref{pSstar} can be rewritten as 
\begin{eqnarray}\label{ss1}
e^{2\lambda}=\frac{1}{y^2},\quad e^{2\nu}=\frac{1}{4}(3y_{1}-y)^2,
\end{eqnarray}
\begin{eqnarray}\label{ss2}
p=\epsilon\left(\frac{y-y_1}{3y_{1}-y}\right),
\end{eqnarray}
\noindent where
\begin{eqnarray}\label{y}
y(r)=\left[1-\left(\frac{r}{\alpha}\right)^2\right]^{1/2},\quad y_1=y(R).
\end{eqnarray}
\noindent We measure the radial coordinate $r$ in the unit
\begin{eqnarray}\label{alpha}
\alpha=\left(\frac{3}{8\pi\epsilon}\right)^{1/2} = \Rs\left(\frac{R}{\Rs}\right)^{3/2}.
\end{eqnarray}
\noindent In terms of \eref{y}, \eref{grr0} and \eref{j} take the form
\begin{eqnarray}
j=\frac{2y}{|3y_1-y|}, \quad\quad \frac{2m(r)}{r}=1-y^2.
\end{eqnarray}
We will find it convenient to introduce the new independent variable \cite{Chandra:1974}
\begin{eqnarray}\label{newx}
x\equiv 1-y=1-\left[1-\left(\frac{r}{\alpha}\right)^2\right]^{1/2},
\end{eqnarray}
\noindent together with the parameter $\ka$ given by
\begin{eqnarray}\label{ka}
\ka \equiv 3\sqrt{1-\frac{R^2}{\alpha^2}} -1 = 3y_{1} - 1.
\end{eqnarray}
In terms of the variables $(\ka,x)$, the HT equations \eref{omegain}, \eref{m0int}, \eref{dp0int}, \eref{v2in} and \eref{h2in} take the following forms
\begin{eqnarray}\label{drag}
x\left(2-x\right)\left(x+\ka\right)\frac{d^2\varpi}{dx^2} + \left[5k + (3-5\ka)x- 4x^2\right]\frac{d\varpi}{dx} -4(\kappa+1)\varpi=0,
\end{eqnarray}
\begin{eqnarray}\label{m0inx}
\frac{dm_0}{dx}=\alpha^3\frac{(1-x)[x(2-x)]^{3/2}}{(\ka+x)^2}\left[\frac{1}{3}x(2-x)\left(\frac{d\varpi}{dx}\right)^2 + \frac{8(\ka+1)}{3(\ka+x)}\varpi^2\right],
\end{eqnarray}
\begin{eqnarray}\label{p0inx}
\frac{1}{\alpha^2}\frac{dp_{0}^{*}}{dx} =&-\frac{\ka+1}{\alpha^2(1-x)(\ka+x)}p_{0}^{*}
-\left[\frac{2+(\ka+1)(1-x)-3(1-x)^2}{\alpha^3(\ka+x)(1-x)^2[x(2-x)]^{3/2}}\right]m_{0}\nonumber\\ 
&+\frac{8x(2-x)}{3(\ka+x)^2}\varpi\left(\frac{d\varpi}{dx}\right) -\frac{8\left[1-(\ka+1)(1-x)\right]}{3(\ka+x)^3}\varpi^2 \nonumber\\
&+\frac{\left[x(2-x)\right]^2}{3(1-x)(\ka+x)^2}\left(\frac{d\varpi}{dx}\right)^2,
\end{eqnarray} 
\begin{eqnarray}\label{v2inx}
\frac{dv_{2}}{dx} =&-\frac{2h_{2}}{\ka+x}
+\frac{2\left[\alpha x(2-x)\right]^2}{3(\ka+x)^3}\left[1+(\ka+1)(1-x)-2(1-x)^2\right]\times\nonumber\\
&\left[\left(\frac{d\varpi}{dx}\right)^2+\frac{4(\ka+1)}{x(2-x)(\ka+x)}\varpi^2\right],
\end{eqnarray} 
\begin{eqnarray}\label{h2inx}
\frac{dh_{2}}{dx} =&\frac{(1-x)^2+(\ka+1)(1-x)-2}{x(2-x)(\ka+x)}h_{2}-\frac{2(\ka+x)}{[x(2-x)]^2}v_{2}\nonumber\\
&+\alpha^2\left[2x^2(2-x)^2-(\ka+x)^2\right]\frac{x(2-x)}{3(\ka+x)^3}\left(\frac{d\varpi}{dx}\right)^2 \nonumber\\
&+4\alpha^2\left[2x^2(2-x)^2+(\ka+x)^2\right]\frac{(\ka+1)}{3(\ka+x)^4}\varpi^2.
\end{eqnarray} 
Equations \eref{drag}-\eref{h2inx} must be supplemented by the relations \eref{h0in}, \eref{m2in} and \eref{p2in} which now take the form
\begin{eqnarray}\label{h0inx}
h_{0}^{-}-\frac{4x(2-x)\alpha^2}{3(\ka+x)^2}\varpi^2 + p_{0}^{*}=h_{c},
\end{eqnarray}
\begin{eqnarray}\label{h2inxx}
h_{2}+\frac{4x(2-x)\alpha^2}{3(\ka+x)^2}\varpi^2 + p_{2}^{*}=0,
\end{eqnarray}
\begin{eqnarray}\label{j2inx}
m_{2} =& \alpha^3 (1-x)^2 \left[x(2-x)\right]^{1/2}\Bigg\{\frac{2x^3 (2-x)^3}{3(\ka+x)^2}\Bigg[\left(\frac{d\varpi}{dx}\right)^2 \nonumber \\
&+ \frac{4(\ka+1)\varpi^2}{x(2-x)(\ka+x)}\Bigg] - \frac{h_{2}}{\alpha^2} \Bigg\}.
\end{eqnarray}
Solutions to equations \eref{drag}-\eref{h2inx} must match at the surface $\Sigma$ with the exterior solutions \eref{omegaout}, \eref{m0out}, \eref{m0RV}, \eref{h2out} and \eref{v2out}, such that 
\begin{eqnarray}\label{wmatch}
\varpi_{1}=\Omega-\frac{2J}{\alpha^3 (1-y_{1}^2)^{3/2}},
\end{eqnarray}
\begin{eqnarray}\label{m0match}
(m_{0})_{1} =&\delta M - \frac{J^2}{\alpha^3 (1-y_{1}^2)^{3/2}}\nonumber \\
&-8\pi{\alpha^3}(1-y_{1}^2)^{3/2}\left(\frac{R}{2M}-1\right)\epsilon(R)(p_{0}^{*})_{1},
\end{eqnarray}
\begin{eqnarray}\label{h0match}
(h_{0})_{1} = - \frac{(m_{0})_{1}}{\alpha\,y_{1}^2(1-y_{1}^2)^{1/2}},
\end{eqnarray}
\begin{eqnarray}\label{h2xout}
(h_{2})_{1} = \frac{3-y_{1}^2}{\alpha^4\,(1-y_{1}^2)^3}J^2 + KQ_{2}^{\;2}(\zeta),
\end{eqnarray}
\begin{eqnarray}\label{v2xout}
(v_{2})_{1} = -\frac{J^2}{\alpha^4\,(1-y_{1}^2)^2} + K\left(\frac{1-y_{1}^2}{y_1}\right)Q_{2}^{\;1}(\zeta),
\end{eqnarray}
where $\zeta=(1+y_{1}^2)/(1-y_{1}^2)$ and the subscript 1 indicates the value of the function at the surface $x_1=1-y_1$. Finally, in terms of the variable $x$, \eref{xi02r} and \eref{ellipr} read now
\begin{eqnarray}\label{xi0_func_x}
\xi_{0}(x)=\frac{\alpha(1-x)(\ka+x)}{[x(2-x)]^{1/2}}p_{0}^{*},
\end{eqnarray}
\begin{eqnarray}\label{xi2_func_x}
\xi_{2}(x)=\frac{\alpha(1-x)(\ka+x)}{[x(2-x)]^{1/2}}p_{2}^{*},
\end{eqnarray}
\begin{eqnarray}\label{ellipx}
\varepsilon(x) = \frac{3(1-x)(\ka+x)}{2x(2-x)}\left[h_{2}(x) + \frac{4x(2-x)\alpha^2}{3(\ka+x)^2}\varpi^2\right] - \frac{3}{2}(v_2 - h_2),
\end{eqnarray}
\noindent where in \eref{ellipx} we have substituted $p_{2}^{*}$ for its value given by \eref{h2inxx}.

\section{Methods and results}
\label{Sect:5}

In this section, we present our results of surface and integral properties for slowly rotating Schwarzschild stars, at second order in the angular frequency $\Omega$. We  numerically integrated the HT structure equations for different values of the parameter $\ten$, in the regime beyond the Buchdahl limit $1 < \ten < 9/8$, by using a standard Runge-Kutta algorithm. In order to show the whole sequence of slowly rotating Schwarzschild stars in adiabatic and quasi-stationary contraction, and also as a consistency check, we include results for stars above the Buchdahl limit, i.e. $R/\Rs \geq 9/8$, which were reported by \cite{Chandra:1974} and confirmed by \cite{Posada:2016xxx}. As another consistency check, we constructed codes that solve the HT equations for the Schwarzschild star in the original $(r,R)$ coordinates using analogous procedures, and the computed values of the surface parameters agree. More information about the relative differences between the parameters computed in $(\kappa,x)$ and $(r,R)$ can be found in \ref{appendix2}. We use dimensionless variables where the units in which the quantities are expressed are given in the corresponding descriptions. In Table~\ref{tab:1} we list the surface properties for some selective values of $\ten$. The main results are presented further in figures~\ref{fig:drag}-\ref{fig:ellip}. 

The regime where the parameter $\ten$ goes beyond the Buchdahl limit, i.e. $1<\ten<9/8$, requires some careful analysis due to the presence of the singular radius $R_{0}$ \eref{r0}. In order to consider this region, \cite{Posada:2016xxx} imposed the condition $\ka=\vert 3y_1-1\vert$ (see equation (48) there); however, as it was pointed out by \cite{Beltracchi:2021lez}, that assumption turned out to be erroneous which marred the results of the surface and integral properties in the regime beyond the Buchdahl limit. However, as we show in the following sections, in the gravastar limit as $R\to\Rs$, the results reported in \cite{Posada:2016xxx} agree with the ones we are presenting here. In the following, we reconsider the analysis carried out by \cite{Posada:2016xxx}, and we study with more detail the behavior of the structure equations at the infinite pressure surface.

Before proceeding further, it is crucial to clarify the range of values of the relevant quantities. We have that in the regime beyond the Buchdahl limit, $1<R/\Rs<9/8$, $\kappa\in (-1,0)$, where $\kappa=0$ corresponds to the Buchdahl radius $\ten=9/8$, and $\kappa=-1$ is the Schwarzschild limit $R=\Rs$. Some relevant values of $x$ are the following
\begin{eqnarray}
\left\{\begin{array}{lr}\label{xvals}
x = 0,  & \mathrm{centre\,\,of\,\,the\,\,star},\nonumber \\
x_{0} = -\kappa, & \mathrm{surface\,\,with\,\,divergent\,\,pressure}, \nonumber \\
x_{1} = (2-\kappa)/3, & \mathrm{surface\,\,of\,\,the\,\,star.} \nonumber
\end{array} \right\}
\end{eqnarray}
Note that in the region beyond the Buchdahl limit, the stellar surface $x_{1}\in(2/3,1)$. Since we are considering this regime here, we have to clarify the region of interest for the integrations. We are primarily concerned with the region with positive pressure, i.e., $x_{0}<x\leq x_{1}$. Regarding the negative pressure region, $0<x<x_{0}$, which emerges once we cross the Buchdahl limit, we will examine the implications of extending the solutions inside the surface $x_0$. The inward extension is not necessary for evaluation of any exterior properties, but it allows for comparison with the interior analytic rotating gravastar solutions.

For configurations with $\ten \geq 9/8$, the analysis is simpler because there is no $x_0$, so the integrations cover the whole interior of the star $0<x<x_1$. The boundary conditions for that situation have already been discussed thoroughly in \cite{Posada:2016xxx,Chandra:1974}, so we will not repeat that here.

\begin{table}[h!]
\center{
\caption{\label{tab:2}Surface properties of slowly rotating Schwarzschild stars, for some selective values of $\ten$. The main results are displayed in the accompanying figures. We employ the same units introduced by \cite{Chandra:1974}, where we include the corresponding factors $c$ and $G$ in case one wishes to recover the physical parameters. The stellar radius is measured in units of the Schwarzschild radius $\Rs\equiv 2GM/c^2$. The normalized moment of inertia $\tilde{I}\equiv I/MR^2$ is dimensionless; the change in mass $\delta M/M$ is measured in the unit $(GJ/\Rs^2 c^3)^2$; the ‘Kerr factor' $\widetilde{Q}\equiv QM/J^2$ is dimensionless; the surface angular velocity $\varpi_{1}$, relative to the local inertial frame, is measured in the unit $GJ/\Rs^3 c^2$; $\xi_0$ ($l=0$ deformation) in the unit $(GJ)^2/\Rs^3 c^6$; $\xi_2$ ($l=2$ deformation) in the unit $(GJ)^2/\Rs^3 c^6$; the ellipticity $\varepsilon$ is measured in units of $(GJ/\Rs^2 c^3)^2$.}
\item[]\begin{tabular}{cccccccc}
\hline\hline
$\ten$ & $\tilde{I}$ & $\delta M/M$ & $\widetilde{Q}$ & $\varpi_{1}$ & $\xi_{0}$ & $\xi_{2}$ & $\varepsilon$ \\
\hline
  1.001 & 0.9980 & 2.0000 & 1.0000 & 0.0059 & -0.0059 & -0.0060 & 5.9880 \\ 
  1.010 & 0.9803 & 2.0000 & 1.0000 & 0.0588 & -0.0593 & -0.0610 & 5.8860 \\
  1.087 & 0.8493 & 2.0471 & 1.0062 & 0.4356 & -0.4255 & -0.6555 & 5.5517 \\
  1.123 & 0.8015 & 2.1667 & 1.0201 & 0.5663 & -0.4758 & -1.0487 & 5.7831 \\ 
  1.125 & 0.7991 & 2.1783 & 1.0212 & 0.5727 & -0.4739 & -1.0731 & 5.8034 \\
  1.150 & 0.7719 & 2.3479 & 1.0381 & 0.6439 & -0.4240 & -1.3941 & 6.0901 \\
  1.490 & 0.6032 & 4.8430 & 1.6111 & 0.8887 & 1.0245 & -6.3423 & 9.8158 \\
\hline\hline
\end{tabular}
\label{tab:1}
}
\end{table}
\subsection{Dragging of the inertial frames and moment of inertia}
\label{Sect:5.1}
The dragging of the inertial frames can be obtained from equation \eref{drag}, which has regular singular points at $x=0$ and $x=-\ka$, in the region of interest. The relevant singularity, beyond the Buchdahl limit, at which we apply boundary conditions is $x_{0}=-\kappa$, and it will be the starting point of our numerical integration.  In \ref{appendix1} we discuss a series solution to \eref{drag}, which was discussed first by~\cite{Beltracchi:2021}, that can be used to seed the numerical integration slightly past the singular point, but only one of the solutions results in finite behavior of the second order perturbation functions at $x_{0}=-\ka$. The appropriate solution of \eref{drag} has the following behavior near $x_0$ [see equation \eref{frob}],
\begin{eqnarray}\label{drag_bc}
\varpi = \varpi_{0}(\ka + x)^2 + \mathcal{O}\left((x+\ka)^3\right),
\end{eqnarray}   
where we are measuring $\varpi$ in units of $\varpi_{0}$, which is arbitrary. We numerically integrated \eref{drag}, with the boundary condition \eref{drag_bc}, starting from $x_0$ (or rather from $x_0+\epsilon$, with a cut-off value $\epsilon\sim 10^{-7}$), up to the stellar surface $x_1$, for various values of the parameter $\ten$ in the regime beyond Buchdahl $1<\ten<9/8$. As a code test, we computed the surface value $\varpi_{1}$ for a star with $\ten=1.12499$, very close to the Buchdahl radius $\Rb=(9/8)\Rs$, and we obtained $\varpi_1=0.57267$, which is in very good agreement with the value reported in \cite{Chandra:1974} at $\Rb$ (see Table 1 there). The agreement between the $(\kappa,x)$ and analogous $(r,R)$ codes is extremely good for the frame dragging, with average error for $\varpi_1$ on the order of $10^{-8}$ (see \ref{appendix2}).

We are following the same conventions as \cite{Chandra:1974} where $\varpi$ is measured in units of $J/\Rs^3$, thus it will be useful to introduce the following quantities
\begin{eqnarray}
\widetilde{\varpi}\equiv\frac{\varpi}{(J/\Rs^3)},\quad \widetilde{\Omega}\equiv \frac{\Omega}{(J/\Rs^3)}.
\end{eqnarray} 
\begin{figure*}[ht]
\centering
\includegraphics[width=0.495\linewidth]{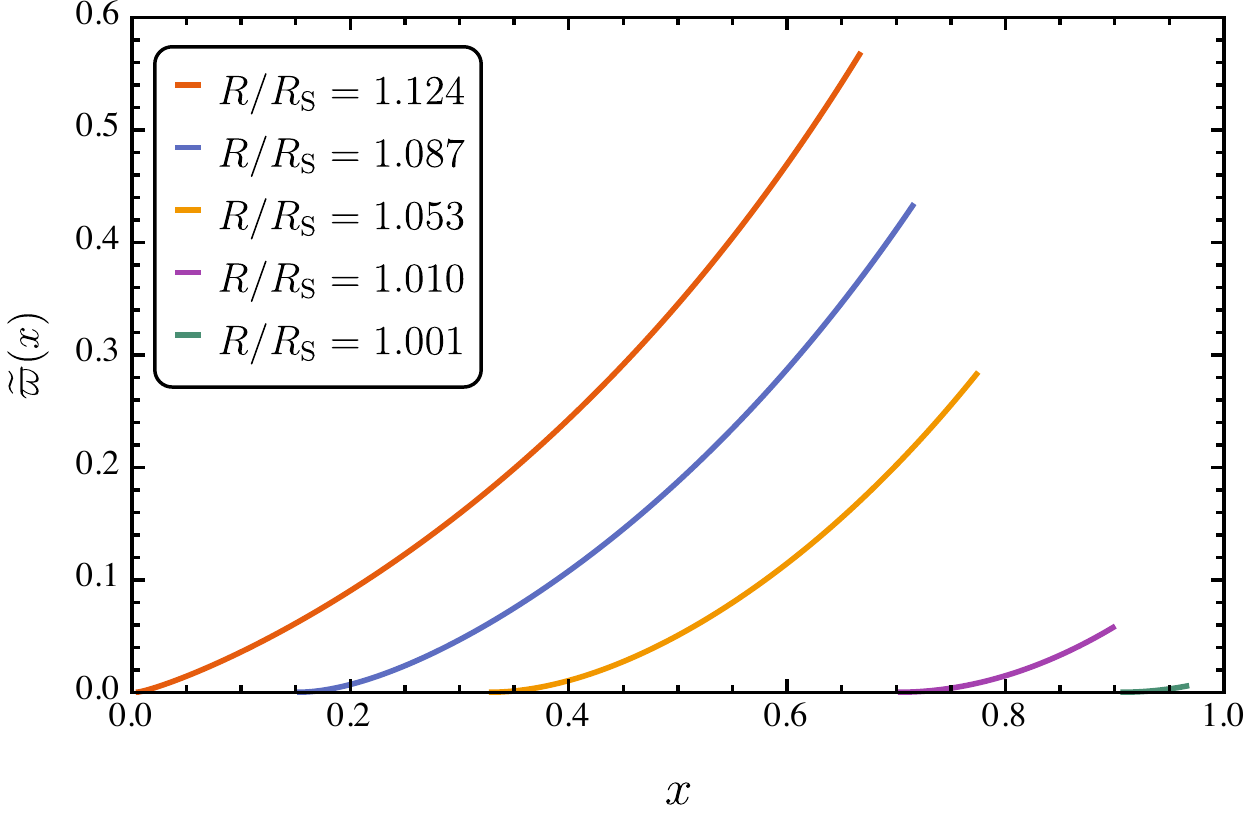}
\includegraphics[width=0.495\linewidth]{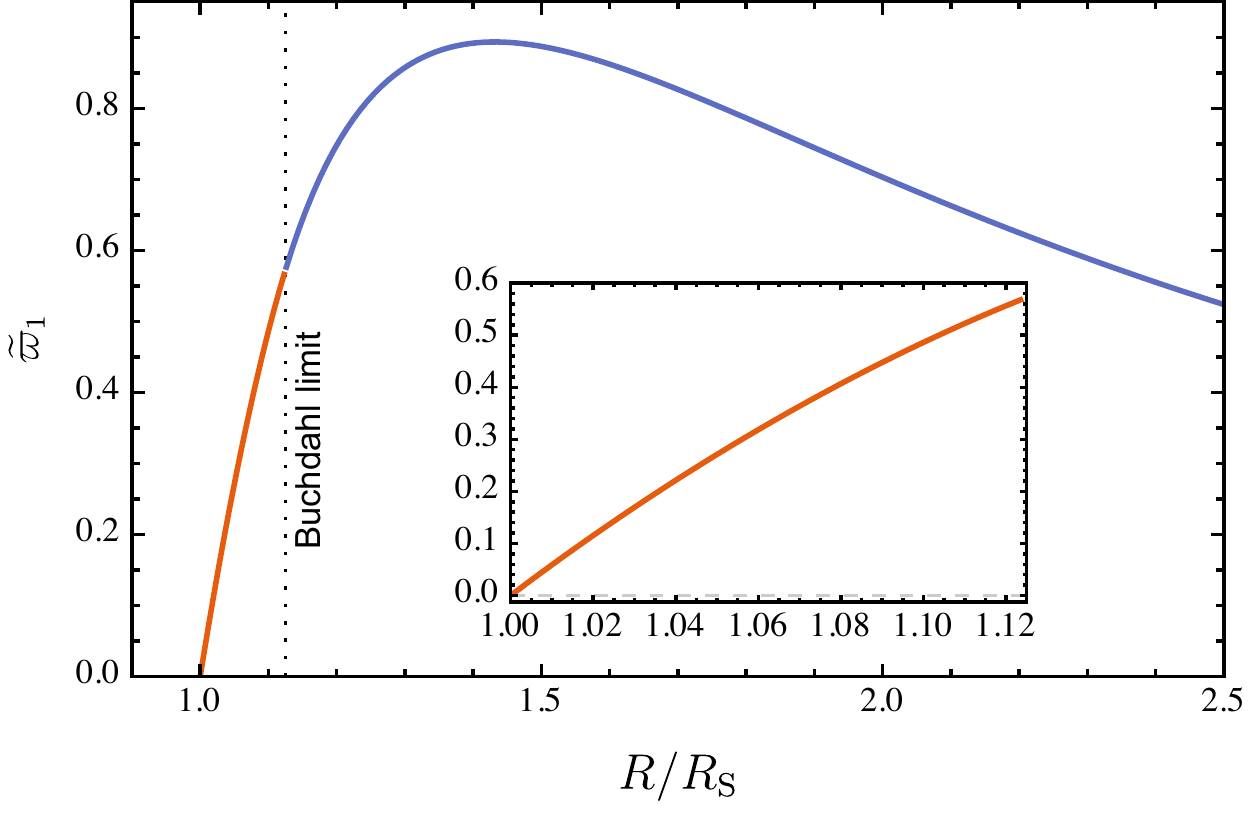}
\caption{{\bf Left panel:} Radial profile of the angular velocity $\widetilde{\varpi}$, relative to the local inertial frame (as measured by a distant observer), for various Schwarzschild stars beyond the Buchdahl limit. Each curve starts at its corresponding $x_0$ value and ends at its corresponding radius $x_1$. {\bf Right panel}: Surface angular velocity $\widetilde{\varpi}_1$, as a function of $\ten$. The inset zooms in the regime beyond the Buchdahl limit $1<\ten<9/8$. Our results for Schwarzschild stars above Buchdahl, i.e. $\ten>9/8$, are in very good agreement with those reported in \cite{Chandra:1974}. Note the smooth match of $\widetilde{\varpi}$ at the Buchdahl radius, and subsequent approach to zero as $R\to \Rs$.}
\label{fig:drag}
\end{figure*}
In the left-hand panel of figure~\ref{fig:drag} we present the radial profiles of the angular velocity $\widetilde{\varpi}$, relative to the local inertial frame, for various values of the parameter $\ten$. Note that each curve starts at its corresponding $x_0$ value and ends at its corresponding radius $x_1$, thus each profile corresponds to the positive pressure region $x_0<x\leq x_1$. We observe that $\widetilde{\varpi}$ is well-behaved and it grows monotonically with $x$ from $x_0$. We also observe how the positive pressure region of the star shrinks as $\ten$ approaches the gravastar limit. In the limit $R\to 2M$, $x_{0}=x_{1}$, then this region practically disappears to become a boundary layer at the Schwarzschild radius, leaving behind a gravastar in the whole interior $x<x_0$ \cite{Mazur:2015kia}. 

In the right-hand panel of figure~\ref{fig:drag} we display the surface value of the angular velocity $\widetilde{\varpi}_1=\widetilde{\varpi}(x_1)$, as a function of the parameter $\ten$. The inset magnifies the region beyond the Buchdahl radius. We observe that $\widetilde{\varpi}_1$ has a local maximum when $R/\Rs\sim 1.4$, it is well-behaved and matches smoothly at the Buchdahl bound. Moreover, in the limit as $R\to \Rs$, $\widetilde{\varpi}_1\to0$; therefore we have that in the gravastar limit
\begin{eqnarray}
\varpi\to 0,\quad \Rightarrow\quad \omega=\Omega,
\end{eqnarray}
\noindent in agreement with \cite{Beltracchi:2021lez}. We point out that $\widetilde{\varpi}_1$ does not reach exactly zero in agreement with a theorem due to Hartle \cite{Hartle:1967he}. For instance, for $\ten=1.000001$, we obtain $\widetilde{\varpi}_1\sim 10^{-6}$. Nevertheless, as it was shown in \cite{Beltracchi:2021lez}, it is expected that in the strict gravastar limit $R=\Rs$, $\widetilde{\varpi}_1=0$, however, we cannot consider exactly that limit in our numerical approach. These results improve those reported by \cite{Posada:2016xxx} where there was a kink at the Buchdahl bound due to the wrong assumption of the absolute value of $\kappa$ (see figure~5 there). However, in the Schwarzschild limit $R\to \Rs$, the results are essentially the same. 

\begin{figure}[ht]
\centering
\includegraphics[width=0.6\linewidth]{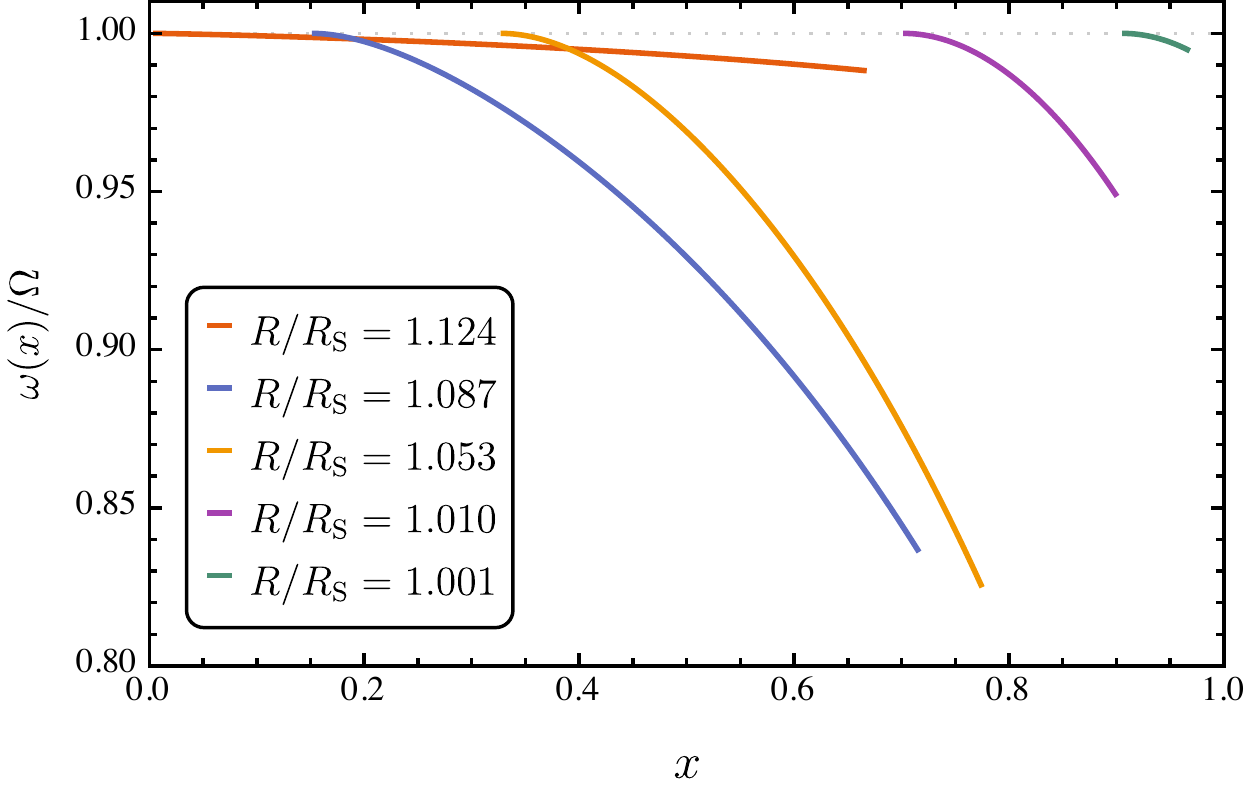}
\caption{Radial profile of the dragging of the inertial frames $\omega(x)/\Omega$, for various Schwarzschild stars in the regime beyond the Buchdahl limit $1<\ten<9/8$. Note that the dragging is maximum at $x_0$, but it is always less than 1 outside $x_0$.}
\label{fig:drag_betas}
\end{figure}

\begin{figure}[ht]
\centering
\includegraphics[width=0.6\linewidth]{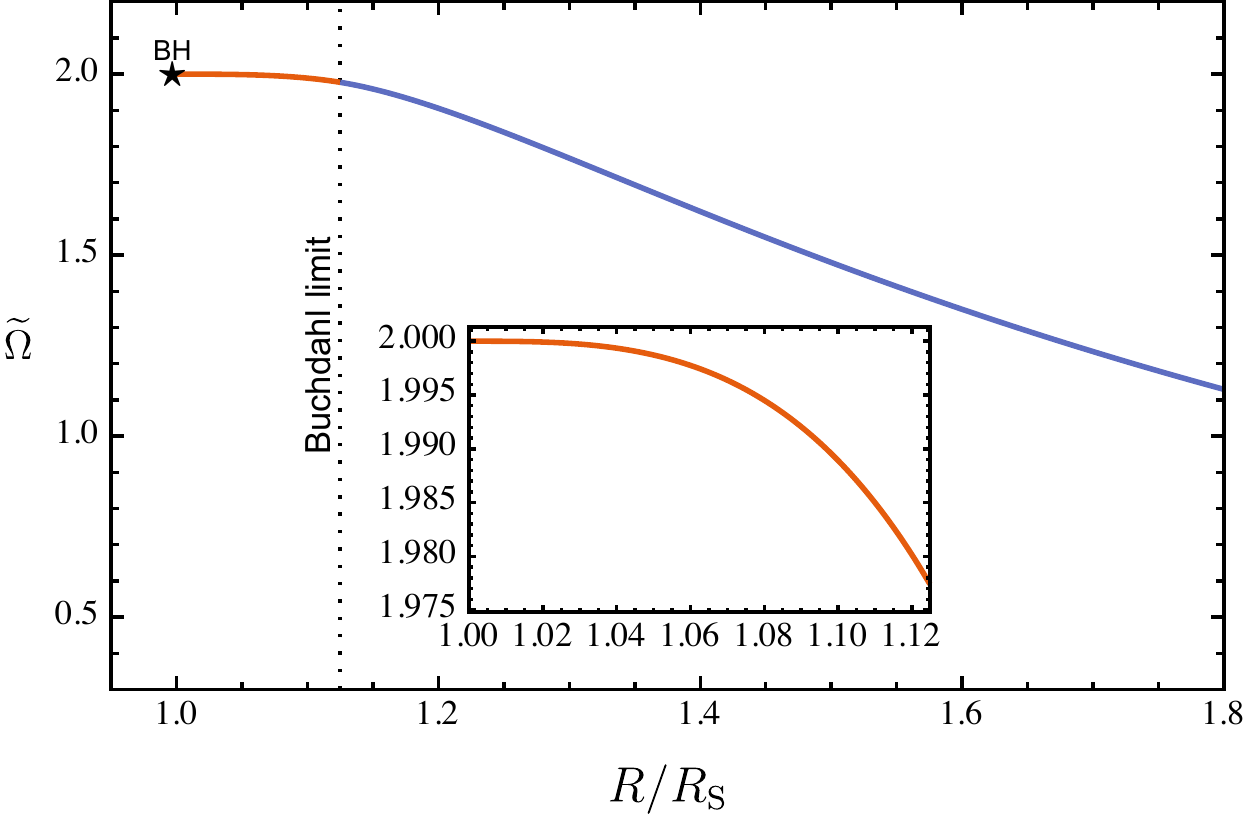}
\caption{Angular velocity of the star $\Omega$ (in units of $J/\Rs^3$), relative to the distant observers, as a function of the parameter $\ten$. The inset enhances the region beyond the Buchdahl limit $1<\ten<9/8$. Note the approach to the value $\Omega=2J/\Rs^3$, as $R\to\Rs$.}
\label{fig:Omega}
\end{figure}

\begin{figure*}
\centering
\includegraphics[width=.495\linewidth]{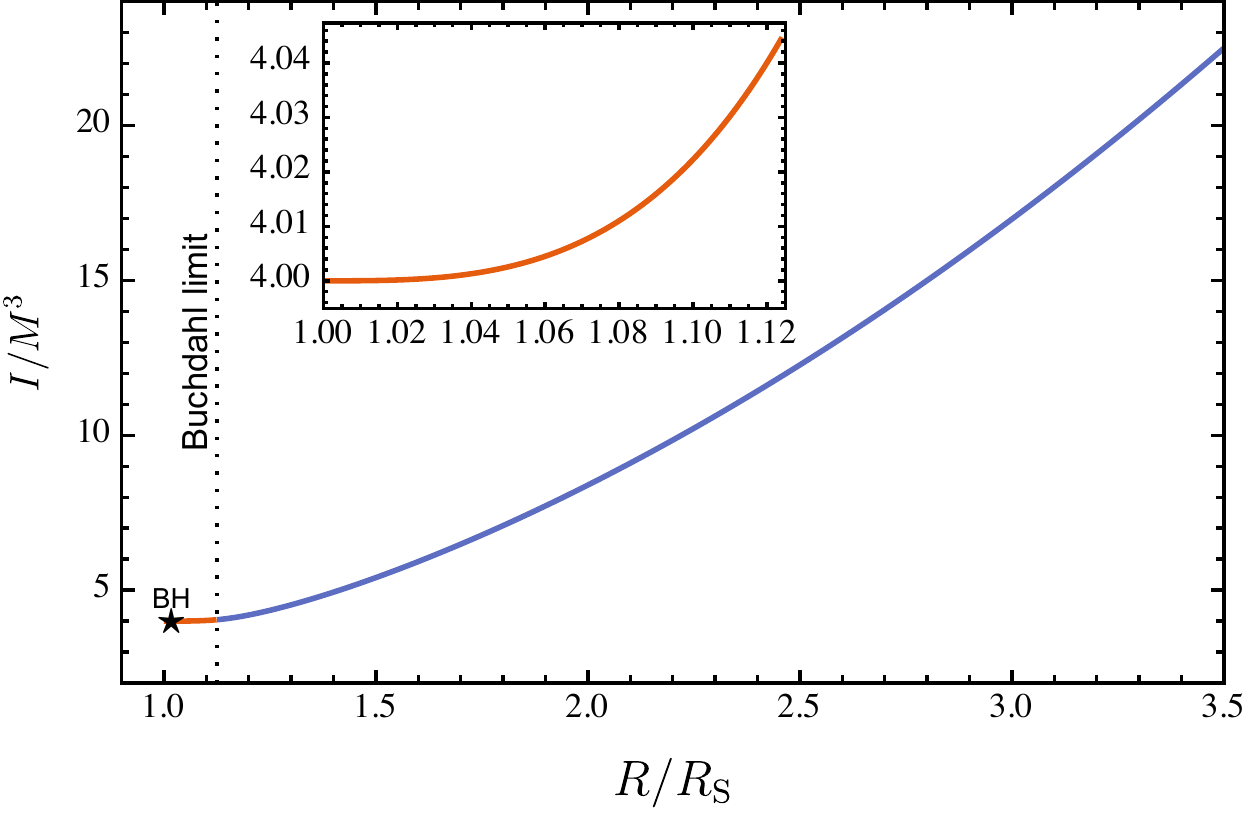}
\includegraphics[width=.495\linewidth]{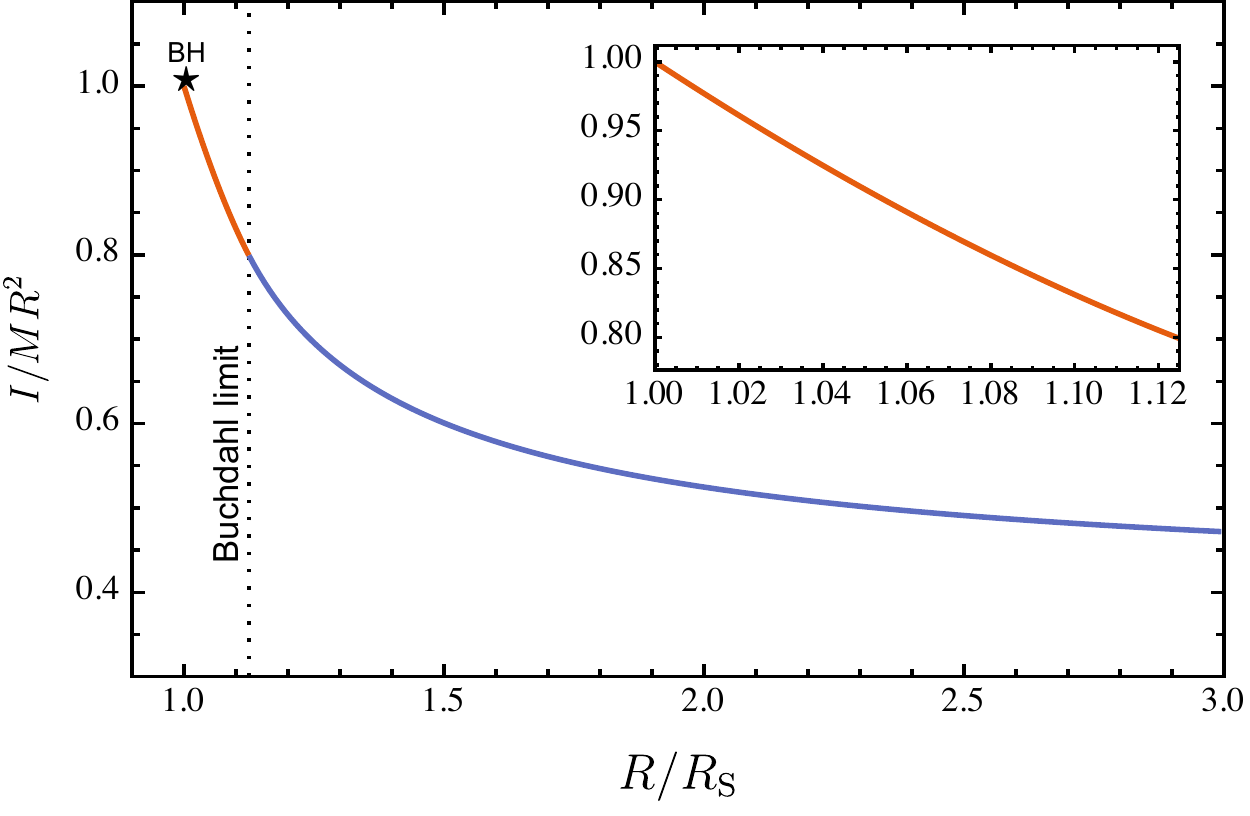}
\caption{\emph{Moment of inertia of Schwarzschild stars}. {\bf Left panel:} Dimensionless moment of inertia $\bar{I}\equiv I/M^3$, as a function of $\ten$. Note the continuous behaviour at the Buchdahl limit, and the subsequent approach to the BH limit $\bar{I}=4$ as $R\to\Rs$. {\bf Right panel:} Same values of $I$ as in the left-hand panel, but now normalized as $\tilde{I}\equiv I/MR^2$, as a function of $\ten$. The inset zooms into the region beyond the Buchdahl limit $\Rs<R<(9/8)\Rs$. Note that $\tilde{I}$ is well-behaved at the Buchdahl limit and it approaches the black hole value $\tilde{I}=1$, as $R\to\Rs$.}
\label{fig:IN}
\end{figure*}

In figure~\ref{fig:drag_betas} we present the radial profiles of the ‘dragging' function $\omega(x)$, normalized by the angular velocity $\Omega$, for some selective values of the parameter $\ten$. Each curve starts (ends) at its corresponding $x_0$ ($x_1$) value. Note that the dragging is maximum at the center, it is 1 exactly on $x_0$ by virtue of the fact that $\varpi\rightarrow0$, and it decreases as we move outwards from $x_0$ up to the surface $x_1$. Thus, the dragging of the inertial frames is maximum at the ‘origin' $x_0$. This is consistent with other results for relativistic stars \cite{Hartle:1968si,Posada:2023bnm}. 
In figure~\ref{fig:Omega} we show the angular velocity of the star $\widetilde{\Omega}$, as a function of the parameter $\ten$. The inset zooms into the region beyond the Buchdahl limit. We observe that $\widetilde{\Omega}$ grows monotonically as the compactness increases, and it is finite and continuous at the Buchdahl bound. Moreover, as $R\to \Rs$ we observe that $\widetilde{\Omega}\to 2$, thus we conclude that in the gravastar limit
\begin{eqnarray}\label{wgr}
\Omega=\omega=\frac{2J}{\Rs^3},
\end{eqnarray}
\noindent which agrees with the exterior HT solution \eref{wsurf}, when $r\to\Rs$. These results improve those we presented in \cite{Posada:2016xxx} (see figure~7 there), where $\widetilde{\Omega}$ went above the value 2, but then it decreased approaching 2 in the gravastar limit. However, as we found before, in the relevant gravastar limit our results here are practically the same as those reported by \cite{Posada:2016xxx}.
The left-hand panel of figure~\ref{fig:IN} shows the dimensionless moment of inertia $\bar{I}\equiv I/M^3$, as a function of the parameter $\ten$. The quantity $\bar{I}$ turns out to be relevant in the context of the $I$-Love-$Q$ relations~\cite{Yagi:2013awa,Yagi:2013bca}. We observe that $\bar{I}$ decreases monotonically as the compactness increases, and it is continuous at the Buchdahl radius. In the gravastar limit, as $R\to\Rs$, $\bar{I}\to4$, corresponding to the BH value. In the right-hand panel of the same figure we display the normalized moment of inertia $\tilde{I}\equiv I/MR^2$, as a function of the parameter $\ten$. We observe that $\tilde{I}$ is finite and continuous at the Buchdahl limit where we obtain $\tilde{I}=0.7992$, which agrees with the value reported in \cite{Chandra:1974}. Moreover, as $R\to \Rs$, $\tilde{I}\to 1$; thus, in the gravastar limit
\begin{eqnarray}\label{inergr}
I=MR^2.
\end{eqnarray}
Notably, this result corresponds to the moment of inertia ascribed to a Kerr black hole. However, for any system satisfying  $p=-\epsilon$,  the `fluid angular velocity' parameter $\Omega$ drops out of the bulk HT equations and becomes ambiguous. As such, the definition of the moment of inertia for Kerr black holes, or for the gravastar in  \cite{Beltracchi:2021lez}, uses the value of frame dragging on the null hypersurface $I=J/\omega_H$ rather than the standard $I=J/\Omega$. In our numerical approach, we can use the standard definition of the moment of inertia and track its limit, which shows how the two definitions agree on the limit.

Note that in our approach, we are contracting the Schwarzschild star \emph{from above}, where $R$ approaches $\Rs$ via an adiabatic and quasi-stationary process. Thus, all the integral and surface properties are computed in the region $x_0<x<x_1$, which in the gravastar limit $R\to\Rs$, it becomes essentially a boundary layer located at the Schwarzschild radius where $x_0=x_1$. Therefore, the moment of inertia \eref{inergr}, is being carried out by this boundary layer, while the interior de Sitter region with negative pressure carries no angular momentum \cite{Beltracchi:2021lez}. Similar to what we found for the dragging function, these results improve those we reported in \cite{Posada:2016xxx}, where we observed a kink at the Buchdahl limit (see figure~9 there). However, in the gravastar limit, the results are the same.

\subsubsection{Inward continuation of the $\varpi$ solution}
 Near $x=0$, the solution for $\varpi$ can be expanded as \cite{Beltracchi:2021}
\begin{eqnarray}
    \varpi=A\sum_{i=0}^{\infty} a_i x^ i+B x^{-\frac{3}{2}} \sum_{i=0}^{\infty} k_i x^ i,
\end{eqnarray}
where $A$ and $B$ are constants and $a_0=k_0=1$.
If we want $\varpi$ to be finite at the origin we must set $B=0$. Doing this and integrating outward typically leads to a nonzero value for $\varpi$ on the infinite pressure surface $x_0$, implying the presence of a $C_2\varpi_2$ term [see equations \eref{omegabarseries2} and \eref{omegabarG}] and giving the higher order perturbation functions divergences \cite{Beltracchi:2021} [see for instance, equation \eref{m0diverge}].

\begin{figure}[ht]
\includegraphics[width=0.6\linewidth]{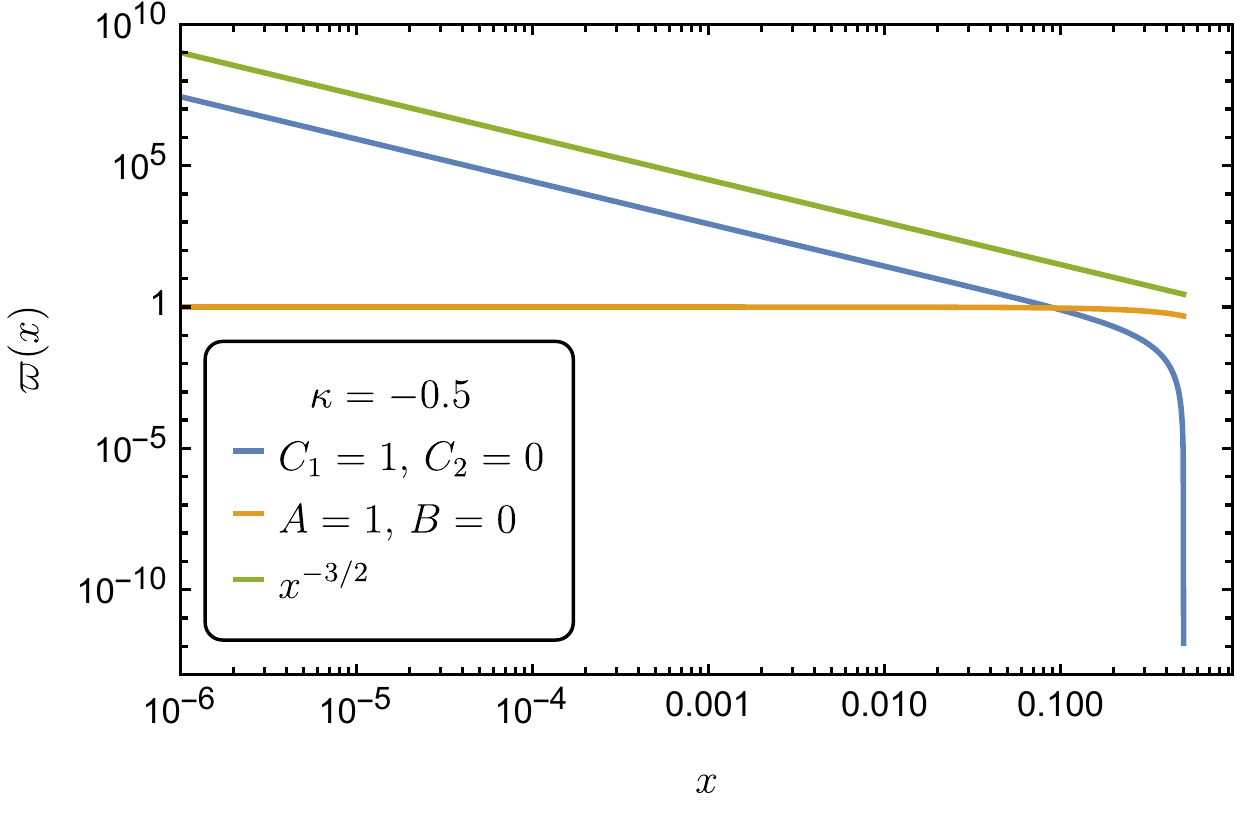}
\caption{ Log-Log plot showing the results of outward integration with $A=1,B=0$ and inward integration with $C_1=1,C_2=0$ for $\kappa=-0.5$. We see that while $A=1,B=0$ is finite at the origin, it is nonzero at the infinite pressure surface which would lead to divergences in the higher order perturbation functions. Conversely, $C_1=1,C_2=0$ is zero on the infinite pressure surface such that the higher order perturbation functions may be finite there, but leads to a $x^{-3/2}$ or equivalently $r^{-3}$ type divergence at the origin. }
\label{fig:win}
\end{figure}

Conversely, integrating $\varpi$ backward from $x_0$  with a nonzero $C_1$, but zero $C_2$, like we use outside the infinite pressure surface, leads to nonzero $B$ and diverging $\varpi$ at the origin. These behaviors are illustrated in figure~\ref{fig:win} for an example value of $\kappa=-0.5$. It is noteworthy that the $x^{-3/2}$ divergence in $\varpi$ with nonzero $B$ corresponds to a $r^{-3}$ divergence, which has the same behavior as a $W_2$ term in the analytic slowly rotating gravastar solution from \cite{Beltracchi:2021lez}.
The solution 
\begin{eqnarray}
\varpi=0,\qquad 0\le x<-\ka,
\end{eqnarray}
is special because it both goes to zero at $x_0$ and does not diverge at the origin. The surface $x_0$  is both a spacetime singularity and a coordinate singularity, manifesting as a singularity in \eref{drag}. Since the Wronskian $\varpi_1 \varpi_2'-\varpi_1'\varpi_2=0$ at $x=-\ka$, the uniqueness theorem is violated there. Therefore, it is possible that there may be different series solutions on each side of the singularity. Following a suggestion by P. Gondolo, we join $\varpi=0$ inside the infinite pressure surface to the $\varpi_1$ series outside the infinite pressure surface. This is a $C1$ match. Therefore we use
\begin{eqnarray}
\varpi=0,\quad 0<x<-\kappa;\qquad \lim_{x\rightarrow-\kappa^+} (\varpi)=\varpi_1\label{propbo},
\end{eqnarray} 
to extend the frame dragging function inside the infinite pressure surface, from which we can attempt to compute the other functions.

\subsection{Spherical deformations of the star}
\label{Sect:5.2}

The spherical deformations of the star can be studied from equations \eref{m0inx} and \eref{p0inx} for the $l=0$ sector.
As we briefly review in \ref{appendix1}, $m_{0}$ is divergent at $x_{0}$ if a non-zero function of $\varpi$ goes into the R.H.S of \eref{m0inx}, i.e. if a nonzero contribution from $\varpi_2$ is present. However, using only the $\varpi_1$ branch, as we have done in the previous subsection \ref{Sect:5.1}, allows for well behaved monopole perturbations from the infinite pressure surface outwards. Thus, in terms of the solution \eref{drag_bc}, equations \eref{m0inx} and \eref{p0inx} can be integrated for $m_0$ and $p_{0}^{*}$, with the following behaviors near $x_0$
\begin{eqnarray}\label{m0_bc}
m_0=\frac{4}{3}(\ka+1)[-\ka(\ka+2)]^{5/2}(\ka+x)\alpha^3 \varpi_{0}^2 + \mathcal{O}\left((x+\ka)^2\right),
\end{eqnarray}
\begin{eqnarray}\label{p0_bc}
p_{0}^{*}=-\frac{\left[2\ka(2+\ka)\right]^2}{3(1+\ka)}(\ka+x)(\alpha\varpi_{0})^2 + \mathcal{O}\left((x+\ka)^2\right).
\end{eqnarray}
We integrated numerically \eref{m0inx} and \eref{p0inx} in the range $x_0<x\leq x_1$, for various values of the parameter $\ten$, with the boundary conditions \eref{m0_bc} and \eref{p0_bc}, which are motivated by the following: from \eref{dmoin} we have $m_0=\mathrm{const.}$, inside the infinite pressure surface. Having it to be nonzero, approaching the surface from outside, either means a point mass at the origin and a continuous $m_0$, or a discontinuous $m_0$ and a $\delta$-function discontinuity in the RHS of \eref{m0int}. In order not to make the notation so cumbersome, it will be useful to introduce the following conventions 
\begin{eqnarray}
\widetilde{h_{0}}\equiv\frac{h_{0}}{(J^2/\Rs^4)},\quad
\widetilde{m_{0}}\equiv\frac{m_{0}}{(J^2/\Rs^3)},\quad \widetilde{p_{0}^{*}}\equiv\frac{p_{0}^{*}}{(J^2/\Rs^4)}.
\end{eqnarray}
\begin{figure*}[ht]
\centering
\includegraphics[width=0.495\linewidth]{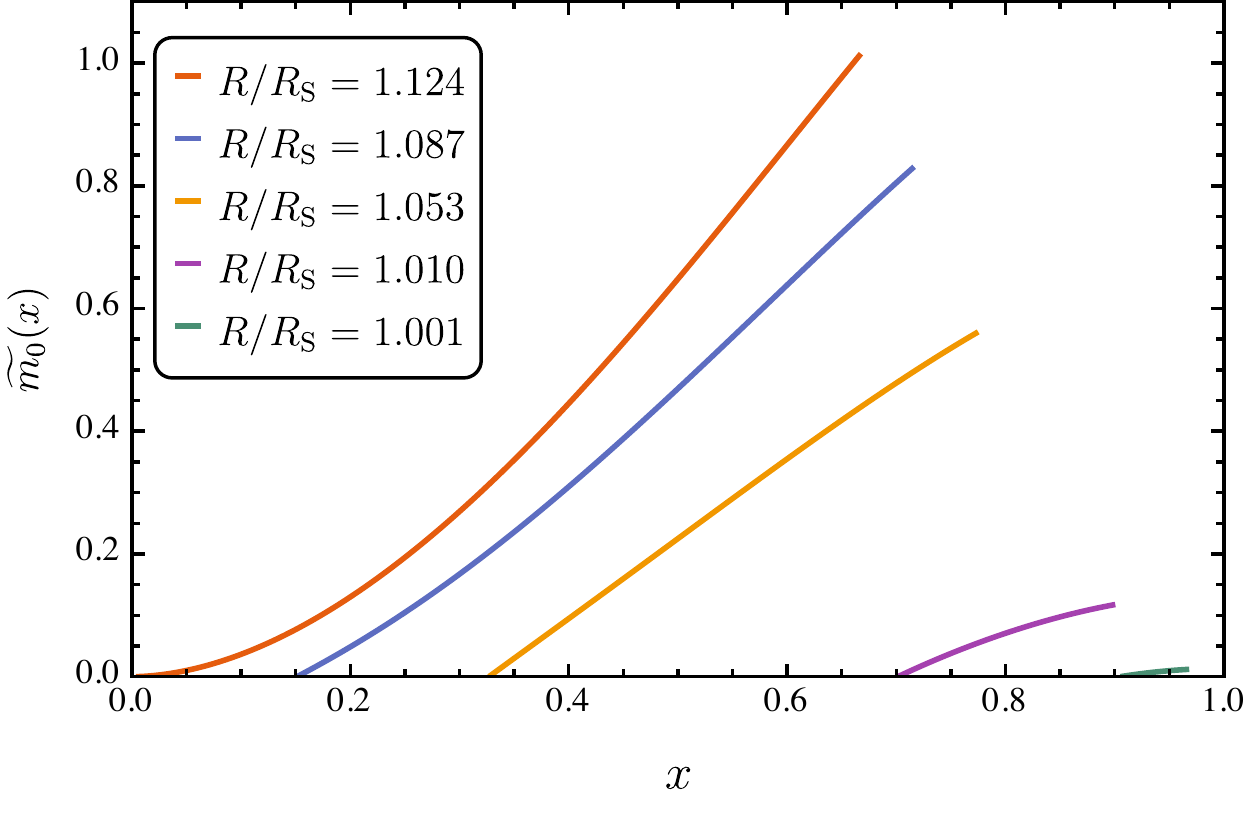}
\includegraphics[width=0.495\linewidth]{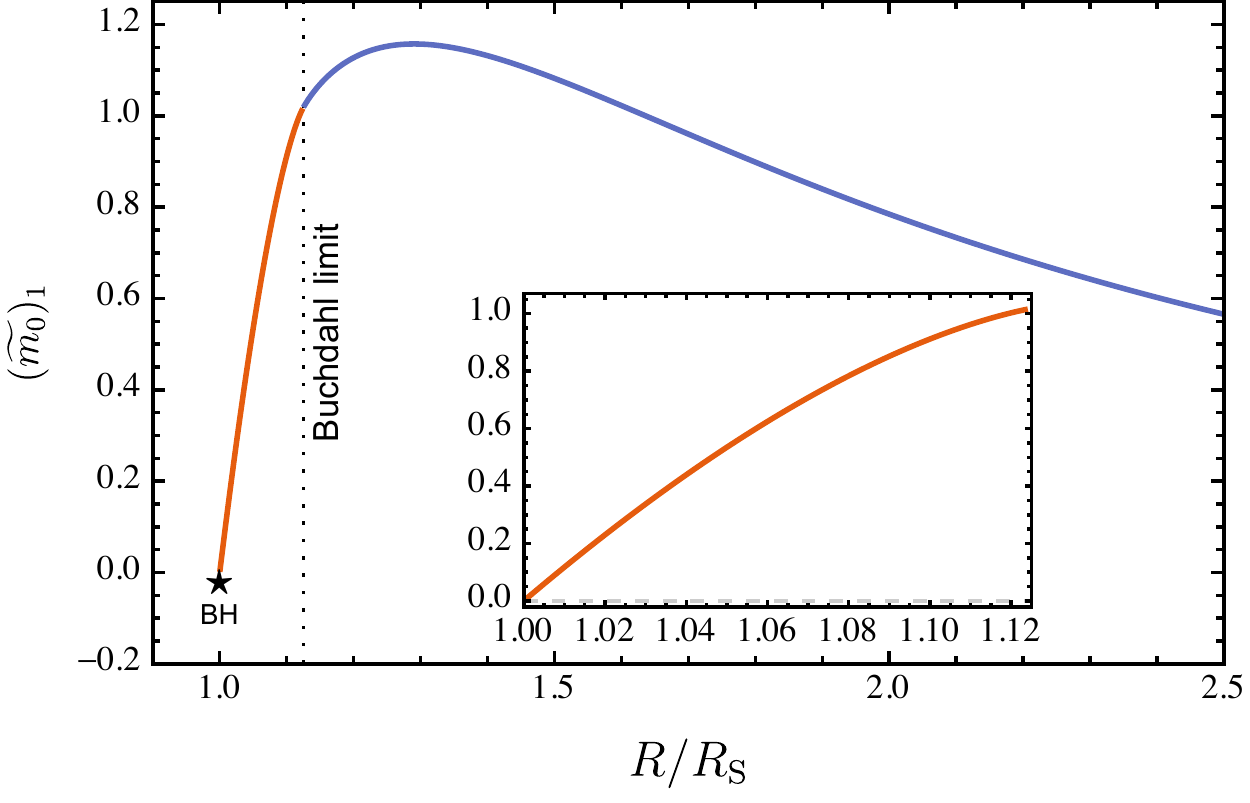}
\caption{{\bf Left panel:} Radial profile of the ‘mass perturbation' function $\widetilde{m_{0}}$, for some selective values of the parameter $\ten$. {\bf Right panel:} Surface value $(\widetilde{m_{0}})_{1}$, as a function of the parameter $\ten$. The inset zooms into the region beyond the Buchdahl limit $1<\ten<(9/8)$. Note that $(\widetilde{m_{0}})_{1}$ has a maximum at $\ten\sim 1.3$. In the gravastar limit $R\to \Rs$, $(\widetilde{m_{0}})_{1}\to0$.}
\label{fig:m0}
\end{figure*}
In the left-hand panel of figure~\ref{fig:m0}, we present the radial profiles of the perturbation function $\widetilde{m_{0}}$, for some selective values of the parameter $\ten$. The various curves start at their corresponding value $x_0$, extending up to their corresponding boundary value $x_{1}$. We observe that $\widetilde{m_{0}}(x)$ is finite and continuous in the whole range $x_{0}<x\leq x_1$. In the right-hand panel of the same figure, we display the value of $\widetilde{m_{0}}$ at the boundary $(\widetilde{m_{0}})_{1}=\widetilde{m_{0}}(1)$, as a function of $\ten$. We include results for configurations above the Buchdahl limit, $\ten>9/8$, where we observe that as $\ten$ approaches the Buchdahl limit (from the right), $(\widetilde{m_{0}})_{1}$ grows monotonically until it reaches a maximum when $\ten\sim 1.3$ in agreement with \cite{Chandra:1974}. We notice that $(\widetilde{m_{0}})_{1}$ is continuous and matches smoothly at the Buchdhal limit; moreover, in the limit $\ten\to1$, $(\widetilde{m_{0}})_{1}\to 0$. Thus, in the gravastar limit 
\begin{eqnarray}\label{j0grav}
m_{0}=0.
\end{eqnarray}
\begin{figure*}[ht]
\centering
\includegraphics[width=0.495\linewidth]{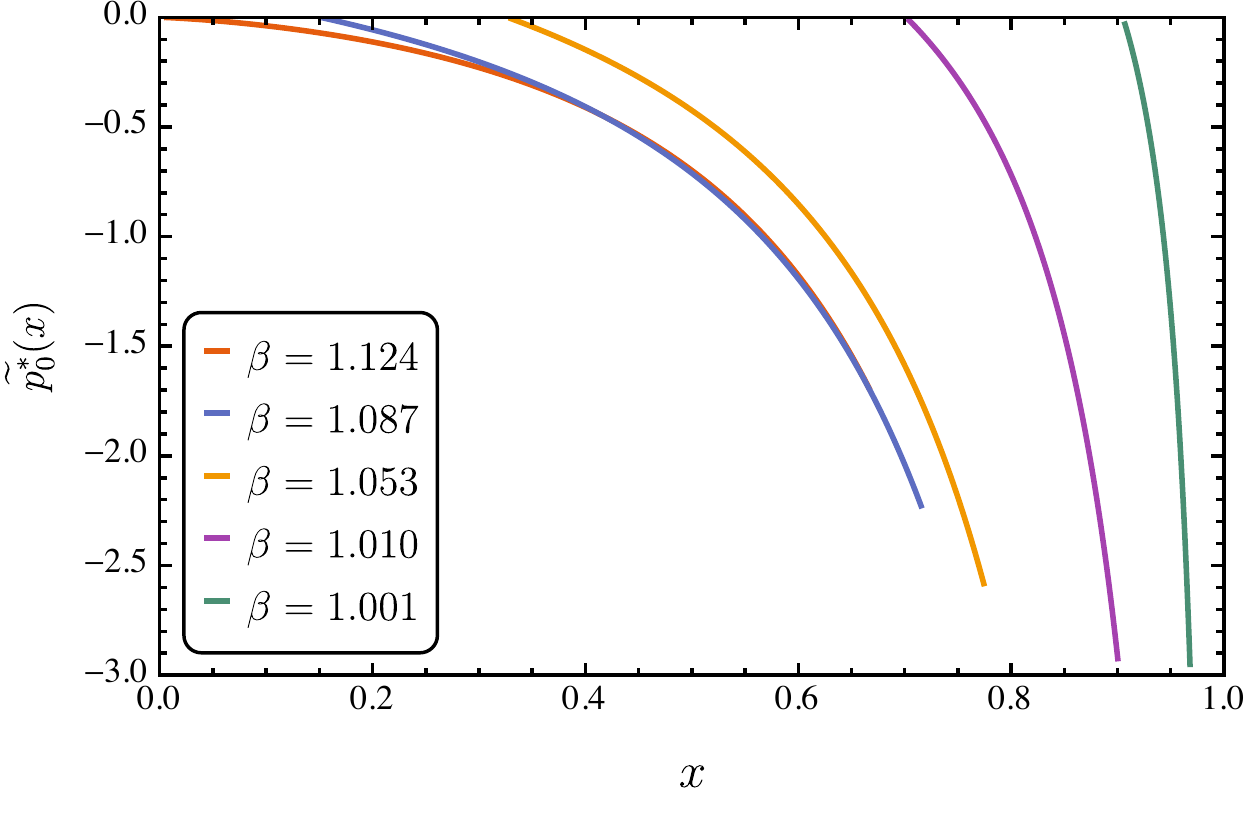}
\includegraphics[width=0.495\linewidth]{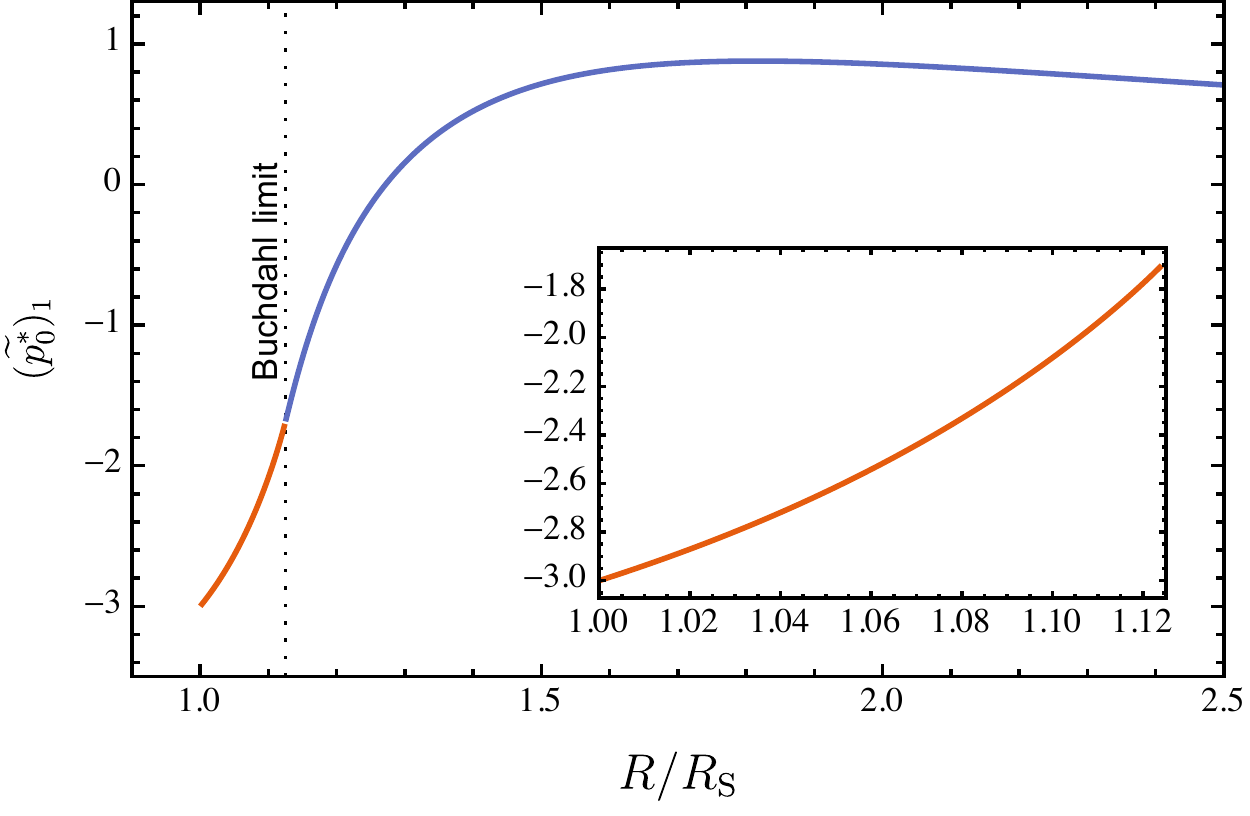}
\caption{{\bf Left panel:} Radial profile of the ‘pressure perturbation' function $\widetilde{p_{0}^{*}}$, for various values of $\ten$. {\bf Right panel:} surface value $\widetilde{p_{0}^{*}}(1)=(\widetilde{p_{0}^{*}})_{1}$, as a function of the parameter $\ten$. Observe that $(\widetilde{p_{0}^{*}})_{1}$ is finite and matches continuously at the Buchdahl limit; in the gravastar limit $\ten\to 1$, $\widetilde{p_{0}^{*}}(1)\to -3$.}
\label{fig:p0}
\end{figure*}
\noindent In the left-hand panel of figure~\ref{fig:p0} we present the radial profiles of the ‘pressure perturbation' function $\widetilde{p_{0}^{*}}$, for various values of $\ten$. Same as before, each curve starts (ends) at its corresponding $x_{0}$ ($x_1$) value. We observe that $\widetilde{p_{0}^{*}}$ is negative, finite, and continuous in the whole corresponding range $x_{0}<x\leq x_{1}$. In the right-hand panel of the same figure, we plot the surface value $(\widetilde{p_{0}^{*}})_{1}=\widetilde{p_{0}^{*}}(1)$, as a function of the parameter $\ten$. We observe that $(\widetilde{p_{0}^{*}})_{1}$ shows a local maximum at $\ten\sim 1.8$, and it goes negative for $\ten< 1.3$ in agreement with \cite{Chandra:1974, Reina:2015jia}.  Note that $(\widetilde{p_{0}^{*}})_{1}$ is finite and continuous across the Buchdahl limit; moreover, in the limit as $\ten\to 1$ we find that $(\widetilde{p_{0}^{*}})_{1}\to -3$. Thus, in the gravastar limit, we have
\begin{eqnarray}
p_{0}^{*}=-\frac{3J^2}{\Rs^4}.
\end{eqnarray}
Having obtained the value of the perturbation function $p_{0}^{*}$ in the gravastar limit, we can determine the constant $h_c$ in \eref{h0inx} which can be rewritten as
\begin{eqnarray}\label{hcx}
h_{c} = h_{0}^{-}(x) - H_{0}(x),
\end{eqnarray}
\noindent where
\begin{eqnarray}\label{H0x}
H_{0}(x) = \frac{4x(2-x)}{3(\ka+x)^2}\alpha^2 \varpi^2 - p_{0}^{*}.
\end{eqnarray}
\noindent At the stellar surface $x=x_1$, or $r=R$, continuity of $h_0$ gives
\begin{eqnarray}\label{hcxR}
h_{c}(R) = h_{0}^{+}(R) - H_{0}(R).
\end{eqnarray}
In figure~\ref{fig:H0_surf} we display the function $H_{0}$, evaluated at the surface, as a function of the parameter $\ten$. We observe that as the compactness increases, $H_{0}$ increases monotonically, until it reaches a local maximum when $\ten=1.065$ (it is noteworthy that the maximum occurs beyond the Buchdahl limit). In the limit as $\ten\to 1$, $H_{0}\to 3$; thus in the gravastar limit we have
\begin{eqnarray}
H_{0}(R)=\frac{3J^2}{\Rs^4}.
\end{eqnarray}
\begin{figure}[ht]
\centering
\includegraphics[width=.6\linewidth]{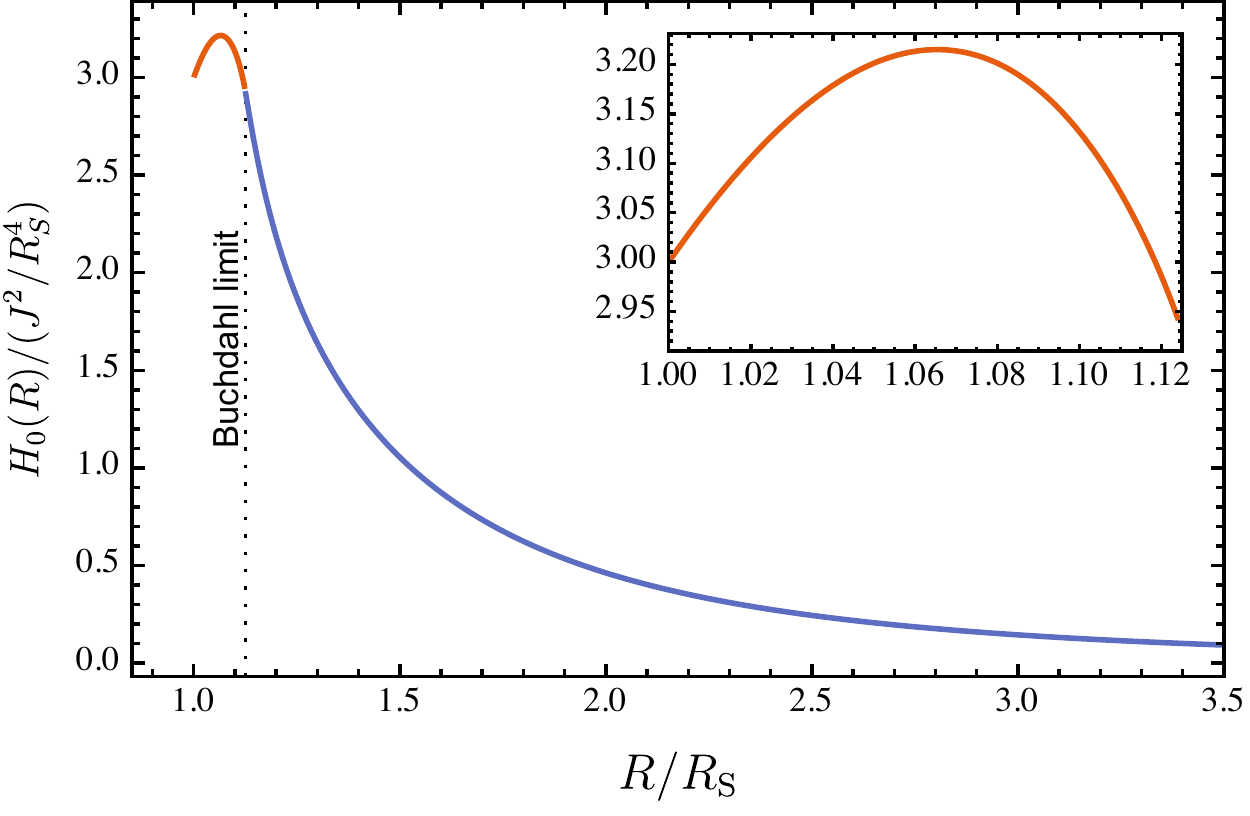}
\caption{Surface value of the function $H_{0}$, as a function of the parameter $\ten$. We measure $H_{0}$ in units of $J^2/\Rs^4$. Observe that in the gravastar limit $\ten\to 1$, $H_{0}(R)\to 3$.}
\label{fig:H0_surf}
\end{figure}
Figure~\ref{fig:hc} shows the profile of $\widetilde{h_{c}}\equiv h_{c}/(J^2/\Rs^4)$, as a function of the parameter $\ten$. We observe that $\widetilde{h_{c}}$ increases negatively as the compactness of the star increases, reaching a local minimum when $\ten\sim 1.195$. In the limit as $\ten\to 1$, $\widetilde{h_{c}}\to -6$, therefore we conclude that in the gravastar limit
\begin{eqnarray}\label{hcgrav}
h_{c}=-\frac{6J^2}{\Rs^4}.
\end{eqnarray}
\begin{figure}[ht]
\centering
\includegraphics[width=.6\linewidth]{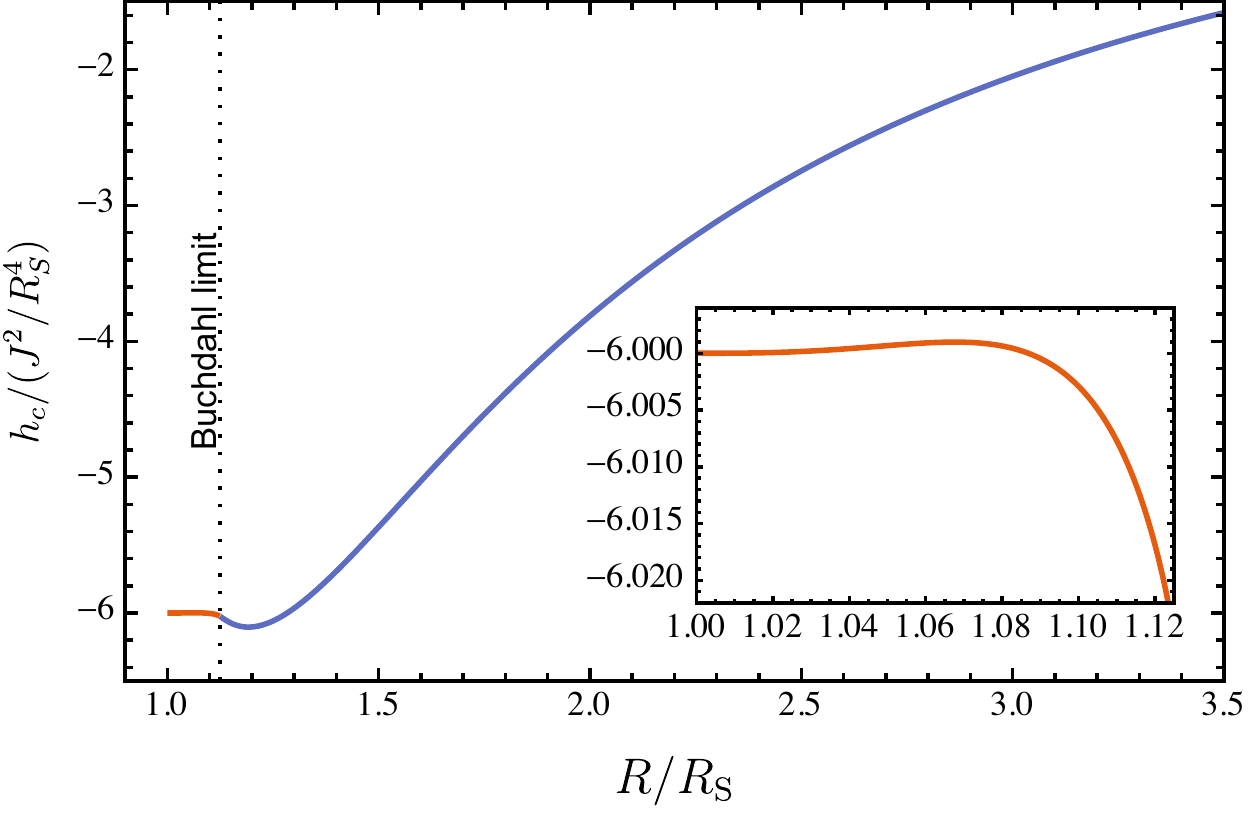}
\caption{Profile of the constant $h_{c}$, as a function of the parameter $\ten$. We measure $h_{c}$ in units of $J^2/\Rs^4$. Observe that in the gravastar limit $\ten\to 1$, $h_{c}(R)\to -6$.}
\label{fig:hc}
\end{figure}
The values of $m_0$ and $p_{0}^{*}$, at the boundary, determine the constant $\delta M$ associated with the change in mass due to the rotation, as given by \eref{m0match}, which in this context can be rewritten as
\begin{eqnarray}\label{deltaMx}
\widetilde{\delta M}=2\left[(\widetilde{m_{0}})_{1} + \left(\frac{R}{\Rs}\right)^{-3}+3\left(\frac{R}{\Rs}-1\right)(\widetilde{p_{0}^{*}})_{1}\right],
\end{eqnarray}
\noindent where we have introduced
\begin{eqnarray}
\widetilde{\delta M}\equiv \frac{\delta M/M}{(J^2/\Rs^4)}.
\end{eqnarray}
\begin{figure}[ht]
\centering
\includegraphics[width=.6\linewidth]{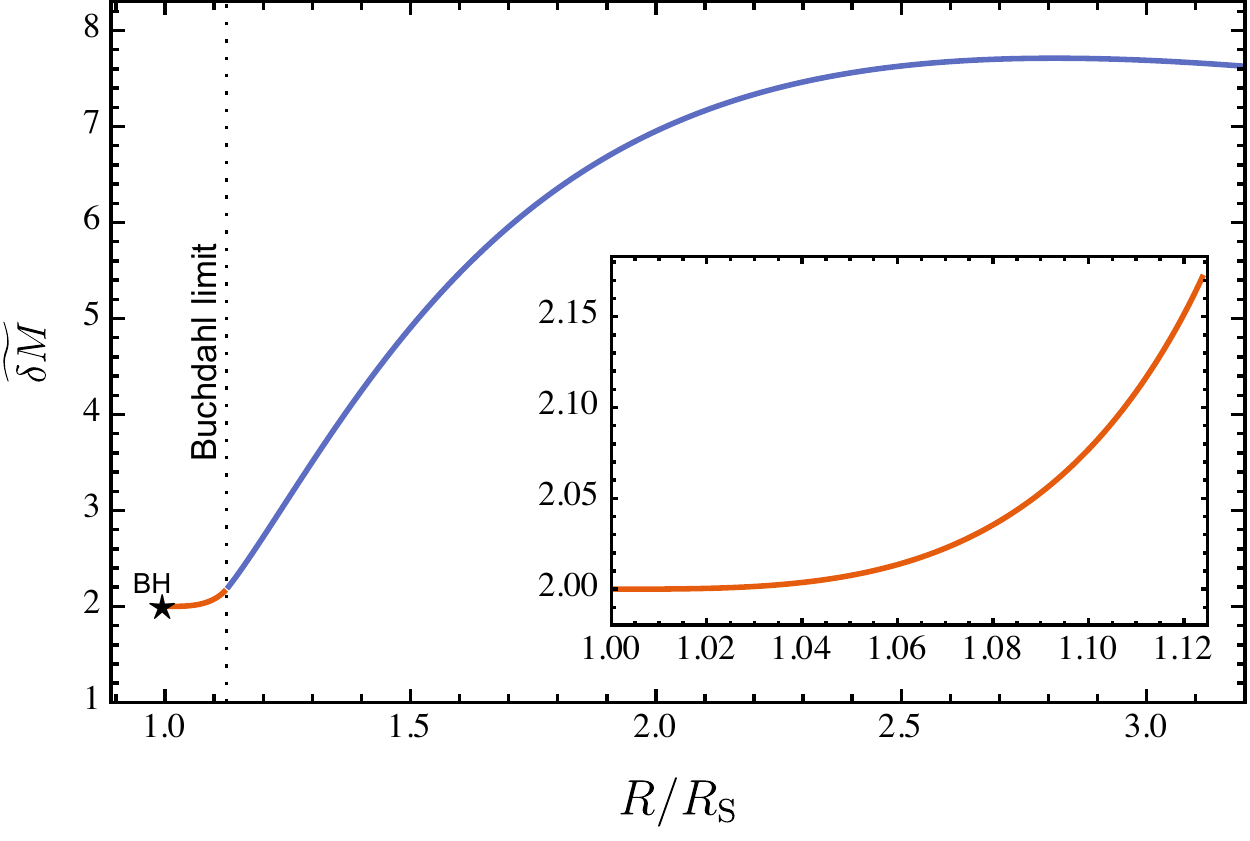}
\caption{Change of mass $\widetilde{\delta M}$, as a function of the parameter $\ten$. The inset zooms into the region beyond the Buchdahl limit $1<\ten<(9/8)$. Note the smooth approach to the value $\widetilde{\delta M}\to 2$, as $R\to\Rs$.}
\label{fig:deltaM}
\end{figure}
In figure~\ref{fig:deltaM} we show our results for $\widetilde{\delta M}$, as a function of the parameter $\ten$. In the inset, we enhance the region beyond the Buchdahl bound, $1<\ten<(9/8)$. We observe that $\widetilde{\delta M}$ has a local maximum when $R/\Rs\sim 2.8$, in agreement with \cite{Reina:2015jia}. Furthermore, $\widetilde{\delta M}$ is well-behaved and continuous at the Buchdahl limit; in the gravastar limit $R\to\Rs$, $\widetilde{\delta M}\to 2$, therefore
\begin{eqnarray}\label{deltaMgrav}
\delta M = \frac{J^2}{\Rs^3}.
\end{eqnarray}
\noindent This result could be already seen from the results of Figs.~\ref{fig:m0} and \ref{fig:p0}, where we note that in the gravastar limit $R\to\Rs$, $(\widetilde{m_{0}})_{1}\to0$; thus, in this limit, the first term of the R.H.S of  \eref{deltaMx} vanishes. Regarding the last term on the R.H.S, although $(\widetilde{p_{0}^{*}})_1$ is non-zero the term in parenthesis goes to zero as $R\to\Rs$, thus we are left only with the second term. 

It is noteworthy that \eref{j0grav} and \eref{deltaMgrav}, for the slowly rotating Schwarzschild star in the gravastar limit, match with the corresponding values associated with the Kerr metric. To see this, we recall that the HT exterior solution, as $r\to \Rs$, takes the following form \cite{Beltracchi:2021lez}
\begin{eqnarray}\label{j0extg}
m_{0} = \delta M - \frac{J^2}{r^3},
\end{eqnarray} 
\begin{eqnarray}\label{h0extg}
h_{0} = -\frac{m_{0}}{r-2M},
\end{eqnarray}
\begin{eqnarray}\label{h2extg}
h_{2} = \frac{J^2}{r^3}\left(\frac{1}{r} + \frac{1}{M}\right),
\end{eqnarray}
\begin{eqnarray}\label{j2extg}
m_{2} = \frac{J^2 (r-2M)(5M-r)}{M r^4},
\end{eqnarray}
\begin{eqnarray}\label{k2extg}
k_{2} = -\frac{J^2 (r+2M)}{M r^4},
\end{eqnarray}
which has been shown to be equivalent to the metric for a slowly rotating Kerr BH with mass $M^{*}=M+\delta M$ and angular momentum $J$, to order $\Omega^2$ \cite{Hartle:1968si}. Note that in the Schwarzschild limit $r\to \Rs$, \eref{j0extg} and \eref{h0extg} agree with \eref{j0grav} and \eref{deltaMgrav}. These results improve those presented in \cite{Posada:2016xxx} (see figure~11 there) where there seems to appear a maximum in the function $\widetilde{\delta M}$, right after crossing the Buchdahl limit, but then $\widetilde{\delta M}$ went below 2, before reaching the corresponding gravastar limit. However, in the gravastar limit $R\to\Rs$, both results are practically in agreement.

\begin{figure}[ht]
\centering
\includegraphics[width=0.6\linewidth]{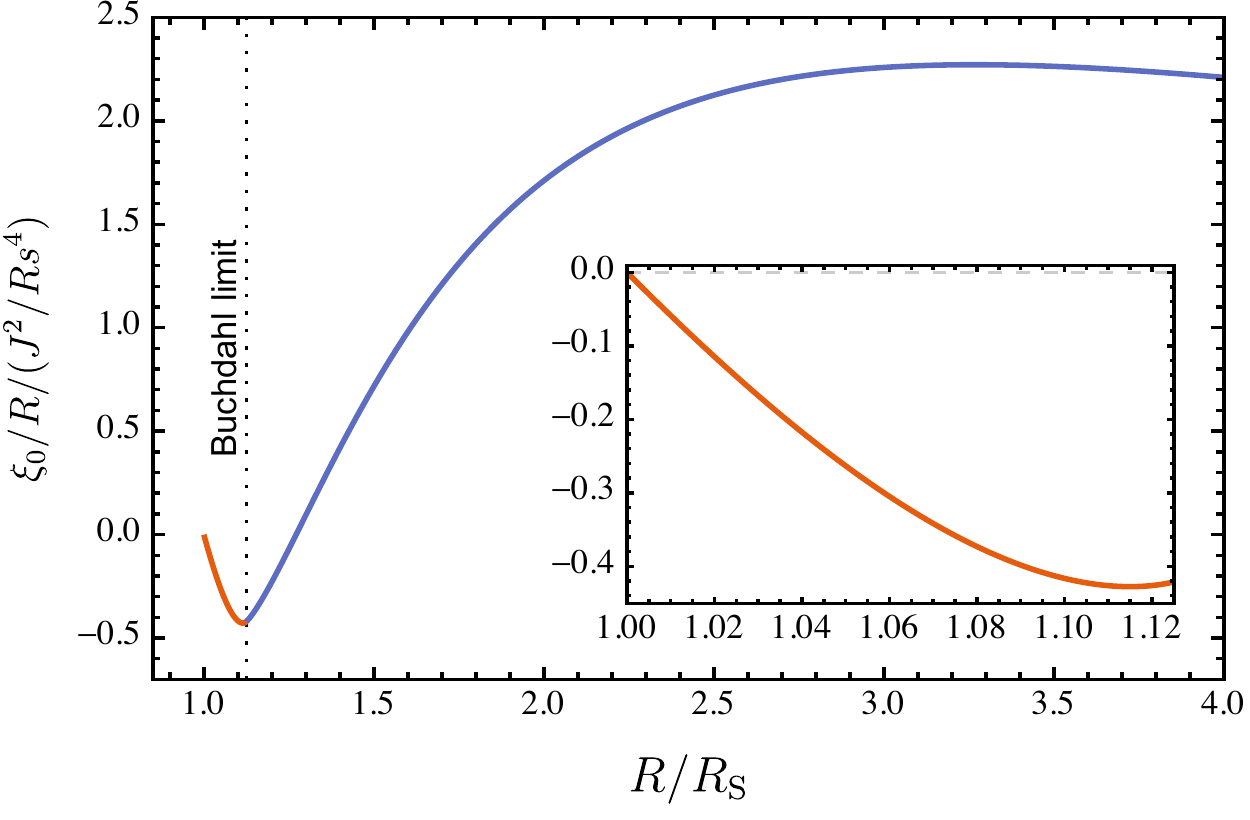}
\caption{The function $\xi_{0}(R)/R$, which describes the $l=0$ deformation of the surface, as a function of the parameter $R/\Rs$. The function $\xi_{0}(R)/R$ is measured in the unit $J^2/\Rs^4$.}
\label{fig:xi0}
\end{figure}

The behaviour of the perturbation function $\xi_{0}(R)/R$, as a function of $R/\Rs$, is shown in figure~\ref{fig:xi0}. First of all, we corroborate the results reported by \cite{Chandra:1974} when $R/\Rs>(9/8)$. We also observe that as the compactness approaches the gravastar limit, $\xi_{0}(R)/R\to 0$. It is interesting that the quantity $\xi_{0}(R)/R$ has a local minimum at $R=1.115$, slightly below the Buchdahl limit.

\subsubsection{Continuation inward of the monopole functions}

Inside the infinite pressure surface $x_0$, using \eref{propbo} into \eref{m0inx} we get
\begin{eqnarray}
\frac{dm_0}{dx}=0\label{dmoin},
\end{eqnarray}
Integration, in this region, trivially leads to $m_0=\mathrm{const.}$ Since we are considering the boundary condition of $\lim_{x\rightarrow-\kappa^+} (m_0)=0$, the most sensible thing to do would be a continuous match\footnote{A discontinuity in $m_0$ is related to a $\delta$-function in the energy density which we should not have here.}, leaving
\begin{eqnarray}
    m_0=0\label{moin},
\end{eqnarray} 
everywhere inside the surface. Thus, using \eref{propbo} and \eref{moin}, equation \eref{p0inx} for $p_0^*$ becomes
\begin{eqnarray}
   \frac{dp_{0}^{*}}{dx} = -\frac{\ka+1}{(1-x)(\ka+x)}p_{0}^{*}, 
\end{eqnarray}
which can be integrated exactly, giving
\begin{eqnarray}
p_{0}^{*} =\frac{c_p (1-x)}{\ka+x}.
\end{eqnarray}
From the algebraic constraint \eref{h0inx} we obtain
\begin{eqnarray}
h_0=h_c-\frac{c_p (1-x)}{\ka+x}.
\end{eqnarray}
Finiteness of the perturbation functions requires $c_p=0$. In many cases, one would also enforce continuity of $h_0$, which would set $h_c$ in the interior, but since $e^{2\nu_0}\rightarrow0$ on the infinite pressure surface any finite discontinuity of $h_0$ will still have a continuous $g_{tt}$. Thus, strict continuity of $h_0$ itself may not be necessary.

\subsection{Quadrupole deformations of the star}
\label{Sect:5.3}

The quadrupole deformations of the star can be studied from equations \eref{v2inx} and \eref{h2inx}, for the $l=2$ sector, whose solutions can be written as a superposition of a particular integral and a homogeneous part (indicated by the superscripts p and h respectively) in the following form 
\begin{eqnarray}\label{superp}
h_{2} = h_{2}^{\mathrm{(p)}} + A h_{2}^{\mathrm{(h)}}, \quad v_{2} = v_{2}^{\mathrm{(p)}} + A v_{2}^{\mathrm{(h)}},
\end{eqnarray} 
\noindent where $A$ is a constant. The homogeneous functions are solutions of the following equations
\begin{eqnarray}\label{v2h}
\frac{dv_{2}^{\mathrm{(h)}}}{dx} = -\frac{2h_{2}^{\mathrm{(h)}}}{\ka+x},
\end{eqnarray}
\begin{eqnarray}\label{h2h}
\frac{dh_{2}^{\mathrm{(h)}}}{dx} = \frac{(1-x)^2 + (\ka+1)(1-x) -2}{x(2-x)(\ka+x)}h_{2}^{\mathrm{(h)}} - \frac{2(\ka+x)}{\left[x(2-x)\right]^2}v_{2}^{\mathrm{(h)}}.
\end{eqnarray}
The following series approximations
\begin{eqnarray}\label{v2h_bc}
    v_2^{\mathrm{(h)}}\approx C+\frac{C (x+\kappa )^2}{8 \kappa ^2+8 \kappa ^3+2 \kappa ^4}+\frac{C (1+\kappa ) (x+\kappa )^3}{\kappa ^3 (2+\kappa
   )^3},
\end{eqnarray}
\begin{eqnarray}\label{h2h_bc}
   h_2^{\mathrm{(h)}}\approx-\frac{C (x+\kappa )^2}{8 \kappa ^2+8 \kappa ^3+2 \kappa ^4}-\frac{3 C (1+\kappa ) (x+\kappa )^3}{2 \kappa ^3 (2+\kappa
   )^3},
\end{eqnarray}
where $C$ is an arbitrary constant, are regular, satisfy the homogeneous equations \eref{v2h} and \eref{h2h}, to order $(x+\kappa)^2$, and can be used for starting the numerical integrations of the homogeneous part of the solution. The particular solutions to equations \eref{v2inx} and \eref{h2inx} are slightly more difficult to deal with because the $\varpi$ dependent driving terms are divergent. The terms are
\begin{eqnarray}
D_{h_2}(\varpi)\approx &\Biggl[-\frac{8\kappa ^3 (2+\kappa )^3}{3 (x+\kappa )}+\frac{8}{3} \kappa^2 (1+\kappa)(2+\kappa)^2\nonumber\\
&-4\kappa(2+\kappa)(x+\kappa)\Biggr]\left(\alpha\varpi_0\right)^2,
\label{dhser}
\end{eqnarray}
\begin{eqnarray}
D_{v_2}(\varpi)\approx&\Biggl[-\frac{8\kappa ^3 (2+\kappa )^3}{3 (x+\kappa )}+\frac{16}{3} \kappa ^2 (1+\kappa ) (2+\kappa)^2\nonumber\\
&-\frac{8}{3}\kappa  (2+\kappa)(x+\kappa)\Biggr]\left(\alpha\varpi_0\right)^2,
\label{dvser}
\end{eqnarray}
to linear order in $x+\kappa$. Obtaining these requires that one keep terms in the series for $\varpi$, at least through $a_2$. While these have divergent $(x+\ka)^{-1}$ terms, it does not mean that $h_2$ or $v_2$ are necessarily divergent. The terms dependent on $h_2$ in  \eref{v2inx}, \eref{h2inx} can also generate divergent $(x+\ka)^{-1}$ terms if $h_2^{\mathrm{(p)}}\rightarrow \mathrm{const.}$, as $x\rightarrow-\ka$. We therefore postulate
\begin{eqnarray}
h_2^{\mathrm{(p)}}\approx \left[h_2^a+h_2^b(x+\kappa)+h_2^c(x+\kappa)^2\right](\alpha\varpi_0)^2\label{h2pser},
\end{eqnarray}
\begin{eqnarray}
v_2^{\mathrm{(p)}}\approx \left[v_2^a+v_2^b(x+\kappa)+v_2^c(x+\kappa)^2\right](\alpha\varpi_0)^2\label{v2pser}.
\end{eqnarray}
We find that for the following values
\begin{eqnarray}
&h_2^a=-\frac{4}{3}\left[\kappa(\kappa +2)\right]^3,\quad &h_2^b=\frac{4}{3} \kappa ^2 (\kappa +1) (\kappa +2)^2,\nonumber\\
&h_2^c=-\frac{2}{3}\ka(2+\ka),\label{h2p_bcf}
\end{eqnarray}
\begin{eqnarray}
v_2^a=0,\quad v_2^b=2 h_2^b,\quad v_2^c=h_2^c,
\label{v2p_bcf}
\end{eqnarray}
$h_2^{\mathrm{(p)}}$ and $v_2^{\mathrm{(p)}}$ satisfy \eref{v2inx} and \eref{h2inx} with error of the order of $(x+k)^2$.

The particular solutions $h_2^{\mathrm{(p)}}$ and $v_2^{\mathrm{(p)}}$ are found by integrating \eref{v2inx} and \eref{h2inx} out from $x_0$ (or rather from $x_0+\delta$, where $\delta=10^{-6}$ is a cut-off value) with the boundary conditions \eref{h2pser} and \eref{v2pser}. Likewise, the homogeneous solutions $h_2^{\mathrm{(h)}}$ and $v_2^{\mathrm{(h)}}$ can be obtained by integrating \eref{v2h} and \eref{h2h} with the initial behaviours \eref{v2h_bc}, \eref{h2h_bc}, for an arbitrary value of the constant $C$. The constant $A$ in \eref{superp} and the constant $K$ appearing in \eref{h2out}, \eref{v2out} and \eref{Q}, can be determined from the matching condition $[h_2]=[v_2]=0$, as given by equations \eref{h2xout} and \eref{v2xout}. 
In the following, will be convenient to introduce the following quantities
\begin{eqnarray}
\widetilde{h_2}\equiv \frac{h_2}{(J^2/\Rs^4)},\quad \widetilde{m_2}\equiv \frac{m_2}{(J^2/\Rs^3)},\quad \widetilde{k_2}\equiv \frac{k_2}{(J^2/\Rs^4)}.
\end{eqnarray} 
\begin{figure*}[ht]
\centering
\includegraphics[width=0.495\linewidth]{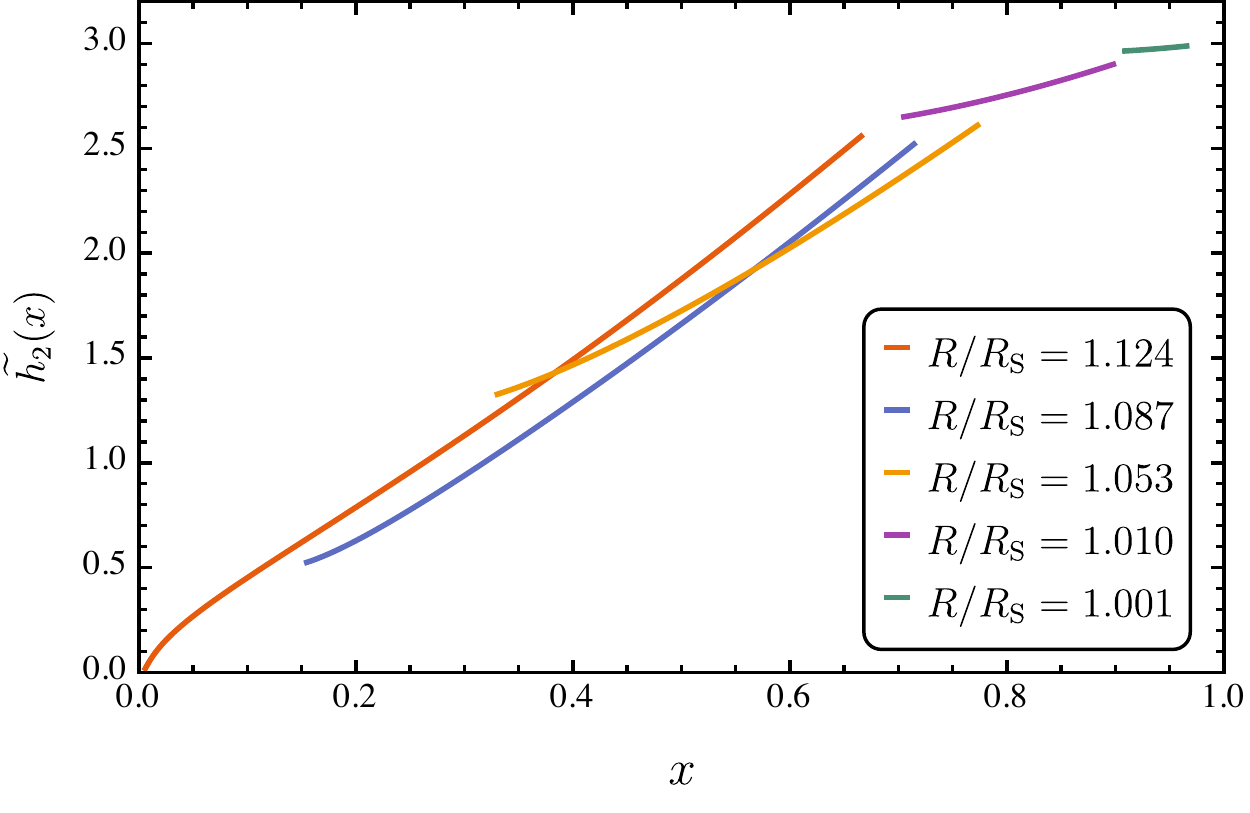}
\includegraphics[width=0.495\linewidth]{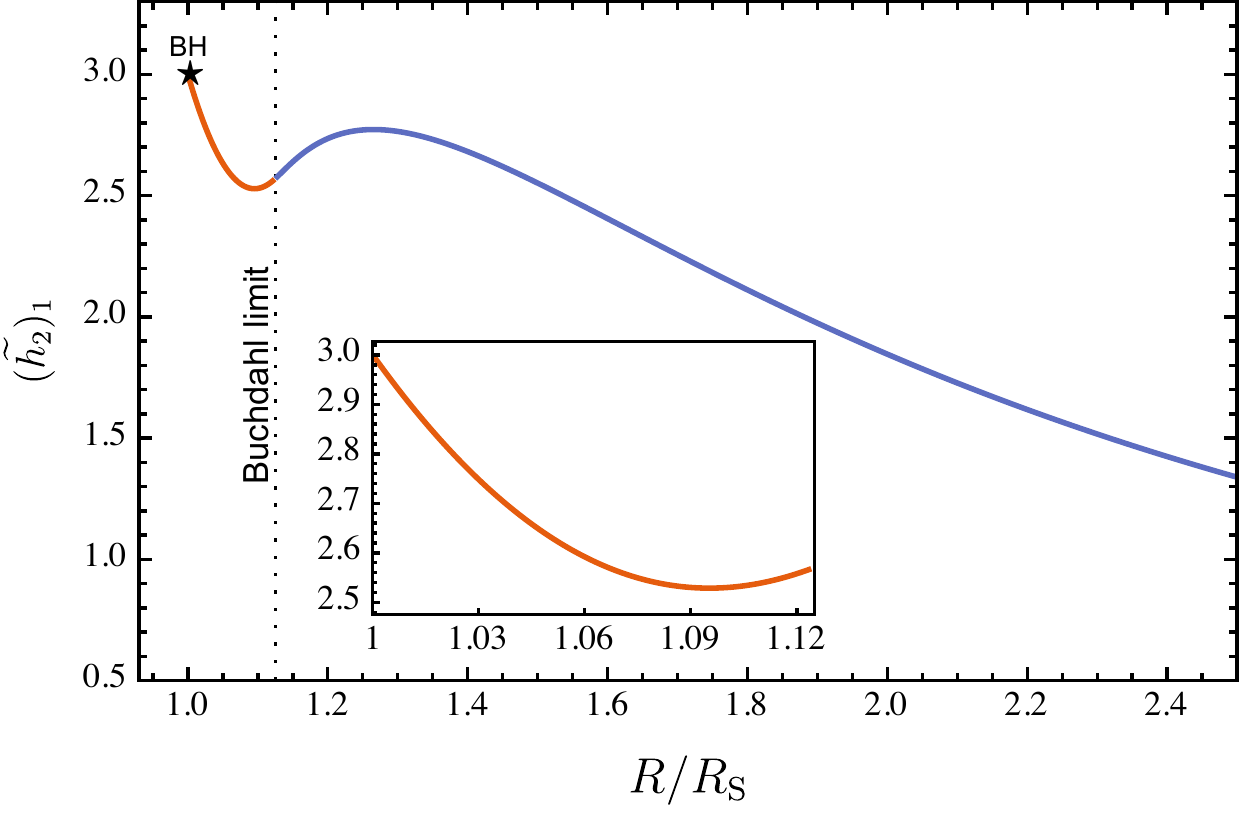}
\caption{{\bf Left panel:} Radial profile of the function $\widetilde{h_{2}}$, for various values of $\ten$. {\bf Right panel:} surface value $(\widetilde{h_{2}})_{1}$, as a function of the parameter $\ten$. The inset zooms into the regime beyond the Buchdahl limit. Observe that $(\widetilde{h_{2}})_{1}$ is finite and matches continuously at the Buchdahl limit; in the gravastar limit $\ten\to 1$, $(\widetilde{h_{2}})_{1}\to 3$.}
\label{fig:h2}
\end{figure*}
In the left-hand panel of figure~\ref{fig:h2} we display the radial profile of the perturbation function $\widetilde{h_2}(x)$, for some selective values of the parameter $\ten$. Likewise to our previous results, each curve starts (ends) at its corresponding $x_0$ $(x_1)$ value. We observe that $\widetilde{h_2}(x)$ is finite and continuous along the corresponding range $x_0<x<x_1$. The right-hand panel of the same figure shows the surface value $(\widetilde{h_2})_{1}=\widetilde{h_2}(1)$, as a function of $\ten$. As before, we also include configurations above the Buchdahl limit, i.e. $R/\Rs > 9/8$. Note that $(\widetilde{h_2})_{1}$ shows a local maximum near $R/\Rs\sim 1.27$ and it matches continuously at the Buchdahl limit; moreover there is a local minimum $\approx2.52$ below the Buchdahl limit at $R/\Rs\sim1.09$, and as $R\to\Rs$, $(\widetilde{h_2})_{1}\to 3$. Thus, we have that in the gravastar limit
\begin{eqnarray}\label{h2_grav}
h_{2}=\frac{3J^2}{\Rs^4}.
\end{eqnarray}
\noindent This result is consistent with the exterior solution for $h_2$, as given by \eref{h2extg}, when $r\to \Rs$. This is expected because the interior and exterior solutions must match continuously at the boundary $\Sigma_{0}=R$, i.e., $[h_2]=0$. 

It is interesting that for the interior of the slowly rotating gravastar, \cite{Beltracchi:2021lez} found that the function $h_2$ takes the form
\begin{eqnarray}
h_2 = \frac{2J^2}{r \Rs}\left(\frac{1}{r^2} - \frac{1}{\Rs^2} \right),
\end{eqnarray} 
which vanishes when $r\to \Rs$. However, the exterior function $h_2$ \eref{h2extg}, evaluated at $r=\Rs$, does not vanish; so it seems that in their model the perturbation $h_2$, besides being non-regular at $r=0$, is discontinuous at the surface. However, the discontinuity in the limit is related to the discontinuity across the infinite pressure surface, which we discuss in the next subsection

\begin{figure*}[ht]
\centering
\includegraphics[width=0.495\linewidth]{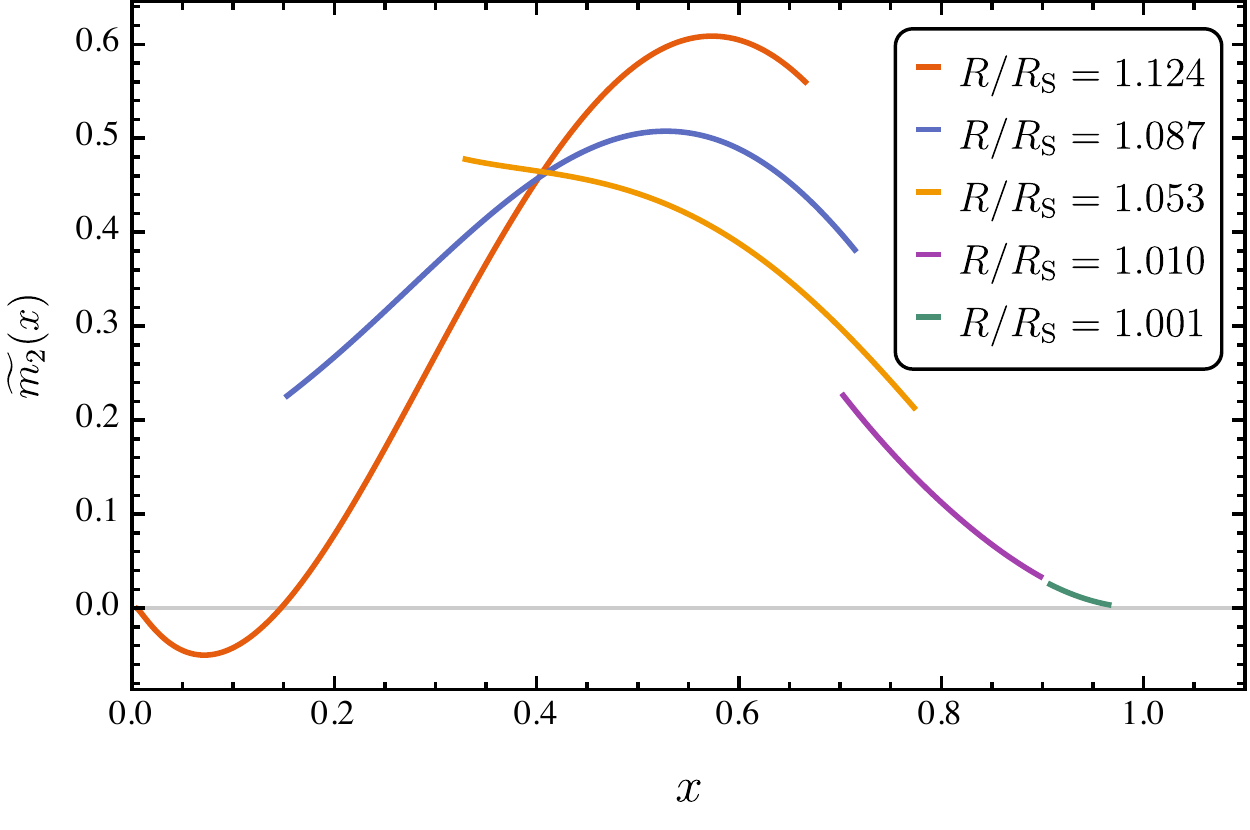}
\includegraphics[width=0.495\linewidth]{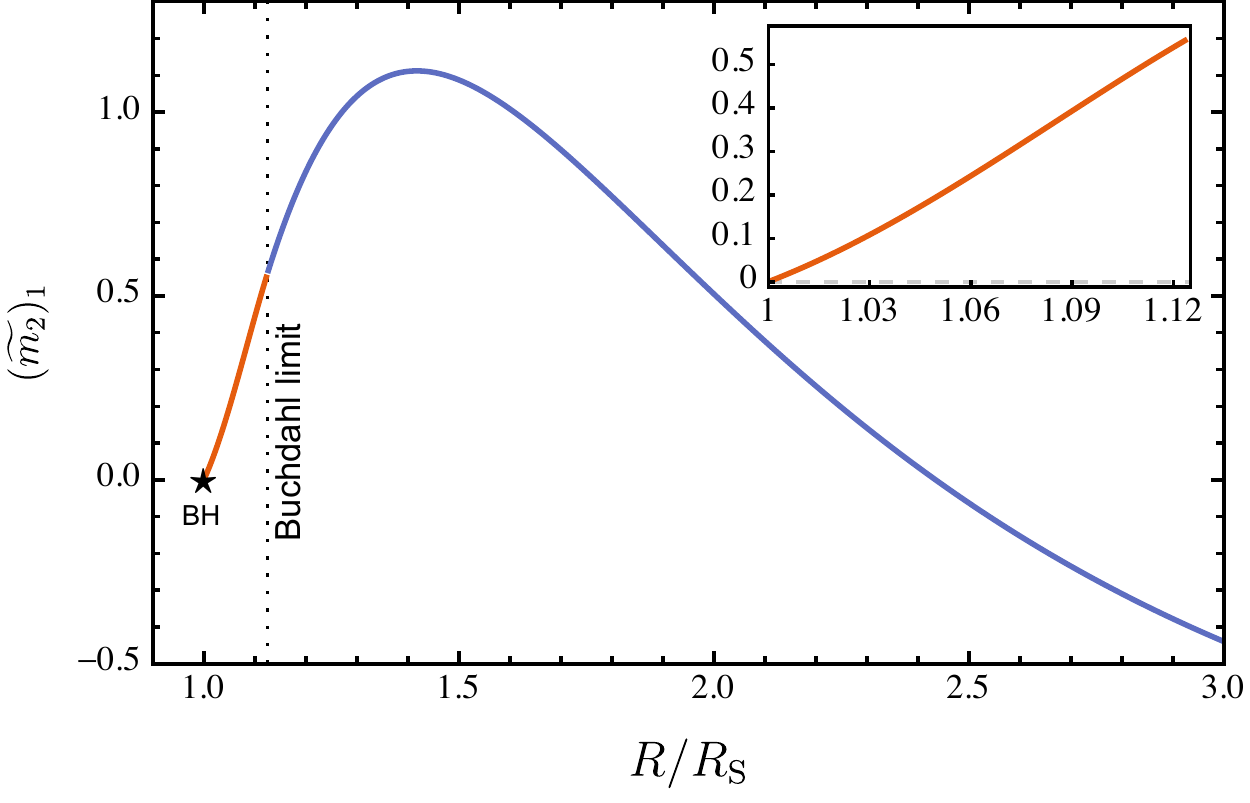}
\caption{{\bf Left panel:} Radial profile of the ‘mass perturbation' function $\widetilde{m_{2}}$ for various values of $\ten$. {\bf Right panel:} surface value $(\widetilde{m_{2}})_{1}=\widetilde{m_{2}}(1)$, as a function of the parameter $\ten$. The inset zooms into the regime beyond the Buchdahl limit. Observe that $(\widetilde{m_{2}})_{1}$ is finite and matches continuously at the Buchdahl limit; in the gravastar limit $\ten\to 1$, $(\widetilde{m_{2}})_{1}\to 0$.}
\label{fig:j2}
\end{figure*}

The perturbation function $\widetilde{m_{2}}(x)$, as given by \eref{j2inx}, is depicted in figure~\ref{fig:j2} where in the left-hand panel we show its radial profile for various values of the parameter $R/\Rs$. In the right-hand panel of the same figure, we display the value at the boundary $(\widetilde{m_{2}})_{1}=\widetilde{m_{2}}(1)$, as a function of the parameter $R/\Rs$.  We include results for configurations above the Buchdahl limit, where we observe that $(\widetilde{m_{2}})_{1}$ has a local maximum when $R/\Rs\sim 1.72$. Close to the Buchdahl limit, $(\widetilde{m_{2}})_{1}$ is finite and continuous, moreover, as $R\to\Rs$ we observe that $(\widetilde{m_{2}})_{1}\to 0$. Thus, we conclude that in the gravastar limit
\begin{eqnarray}\label{j2extgr}
m_{2}=0,
\end{eqnarray}
which agrees with the exterior HT solution for $m_2$, equation \eref{j2extg}, when $r\to \Rs$. As before, this is expected from the continuity condition at the surface $[m_2]=0$. Note that our result also agrees with the surface value of ~(9.1d) in \cite{Beltracchi:2021lez}, when $r\to R_{H}=\Rs$, for the interior of the slowly rotating gravastar. 

\begin{figure*}[ht]
\centering
\includegraphics[width=0.495\linewidth]{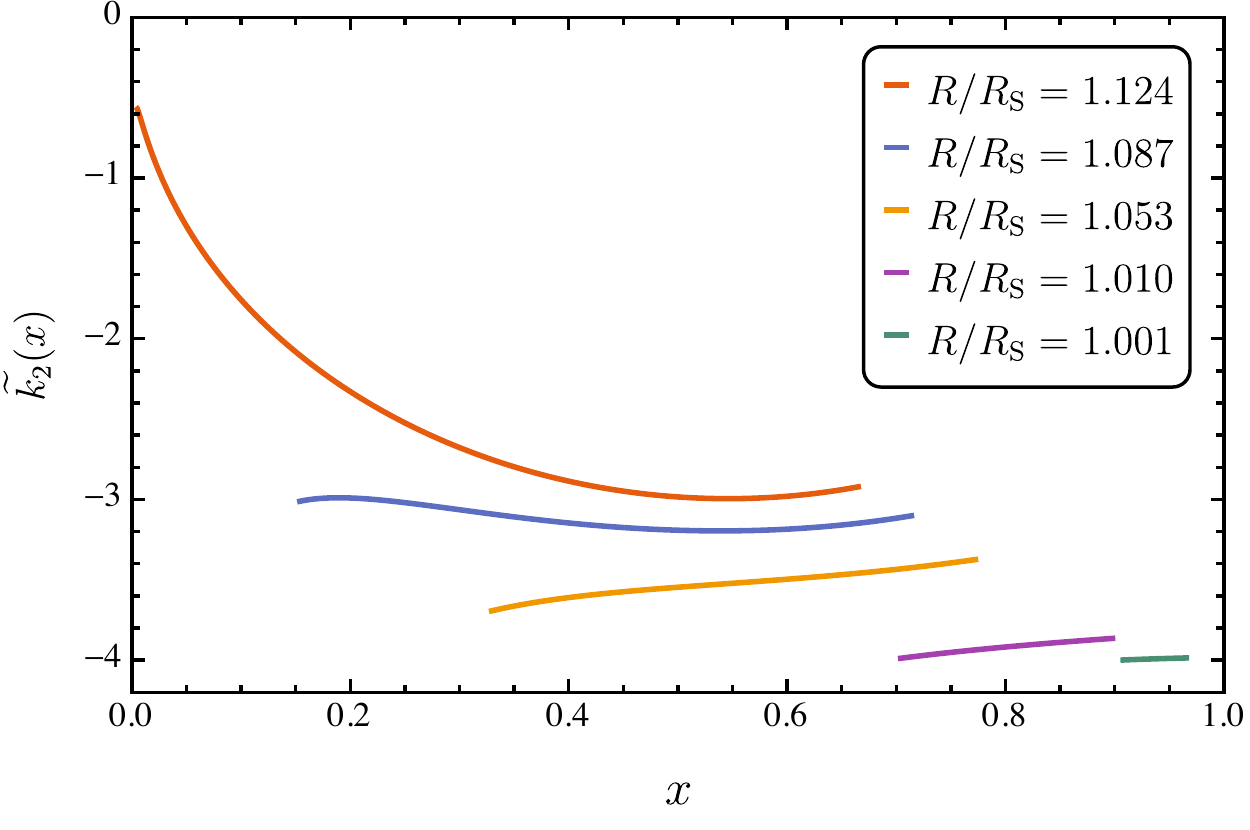}
\includegraphics[width=0.495\linewidth]{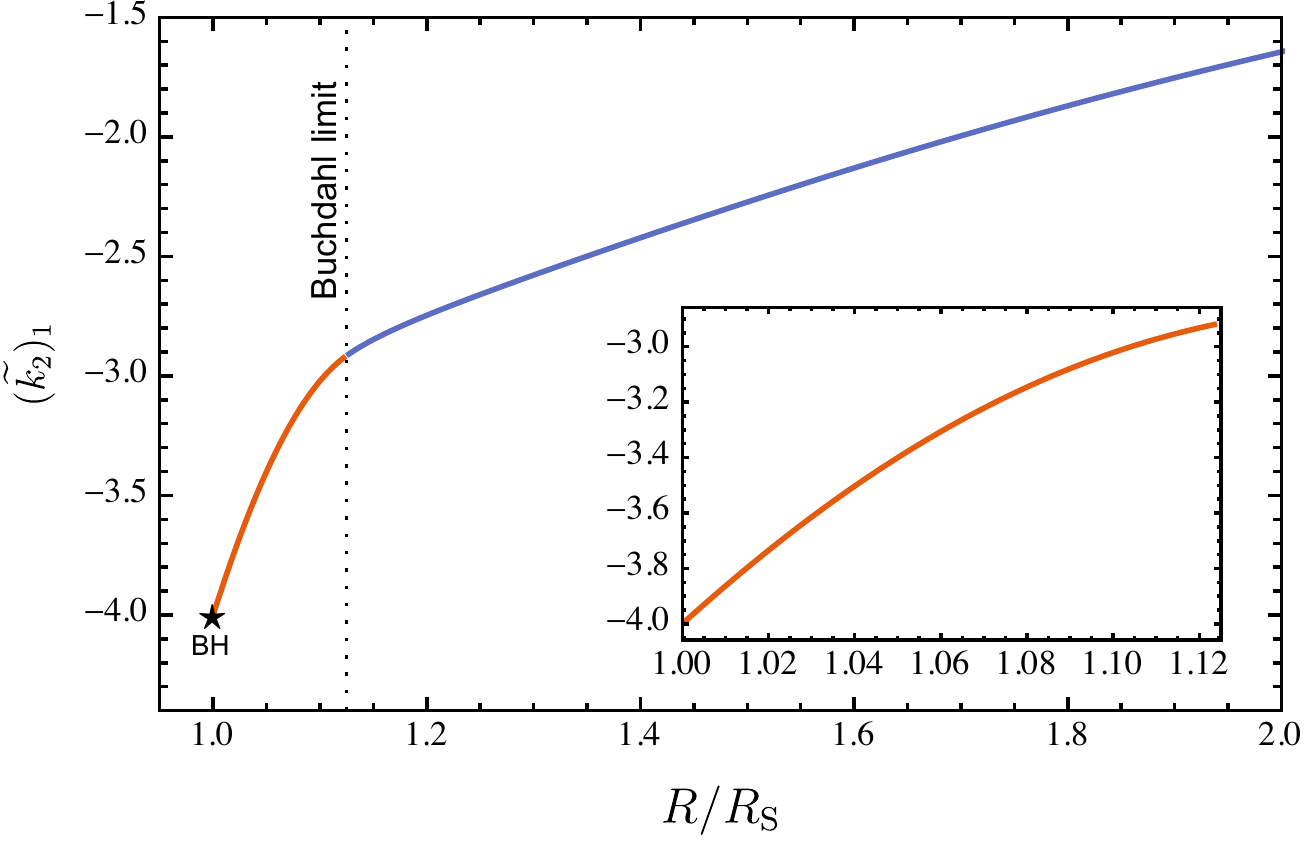}
\caption{{\bf Left panel:} Radial profile of the perturbation function $\widetilde{k_{2}}$, for various values of $\ten$. {\bf Right panel:} surface value $\widetilde{k_{2}}(1)=(\widetilde{k_{2}})_{1}$, as a function of the parameter $\ten$. Observe that $(\widetilde{k_{2}})_{1}$ is finite and matches continuously at the Buchdahl limit; in the gravastar limit $\ten\to 1$, $(\widetilde{k_{2}})_{1}\to -4$.}
\label{fig:k2}
\end{figure*}

The left-hand panel of figure~\ref{fig:k2} shows the radial profile of the perturbation function $\widetilde{k_{2}}(x)$, for different values of the parameter $R/\Rs$. We observe that $\widetilde{k_{2}}(x)$ is finite and continuous in the corresponding regions $x_0<x<x_1$. In the right-hand panel of the same figure, we display the surface value $\widetilde{k_{2}}(1)=(\widetilde{k_{2}})_{1}$, as a function of $R/\Rs$. We observe that $(\widetilde{k_{2}})_{1}$ is well-behaved and matches continuously at the Buchdahl limit; furthermore, in the limit as $R\to\Rs$, $(\widetilde{k_{2}})_{1}\to -4$. Thus, we conclude that in the gravastar limit
\begin{eqnarray}\label{k2extgr}
k_{2}=-\frac{4J^2}{\Rs^4},
\end{eqnarray}
\noindent which agrees with the exterior HT solution, \eref{k2extg}, when $r\to\Rs$. Our result \eref{k2extgr} is also in agreement with the surface value for $k_2$ for the slowly rotating gravastar, equation (9.1f) in \cite{Beltracchi:2021lez}, when $r\to R_{H}=R_{S}$. However, we note that their perturbation function is non-regular as $r\to 0$, which seems to be related to the behavior we find after extending the $k_2$ function inward as we discuss in the next subsection.

\begin{figure}[ht]
\centering
\includegraphics[width=.6\linewidth]{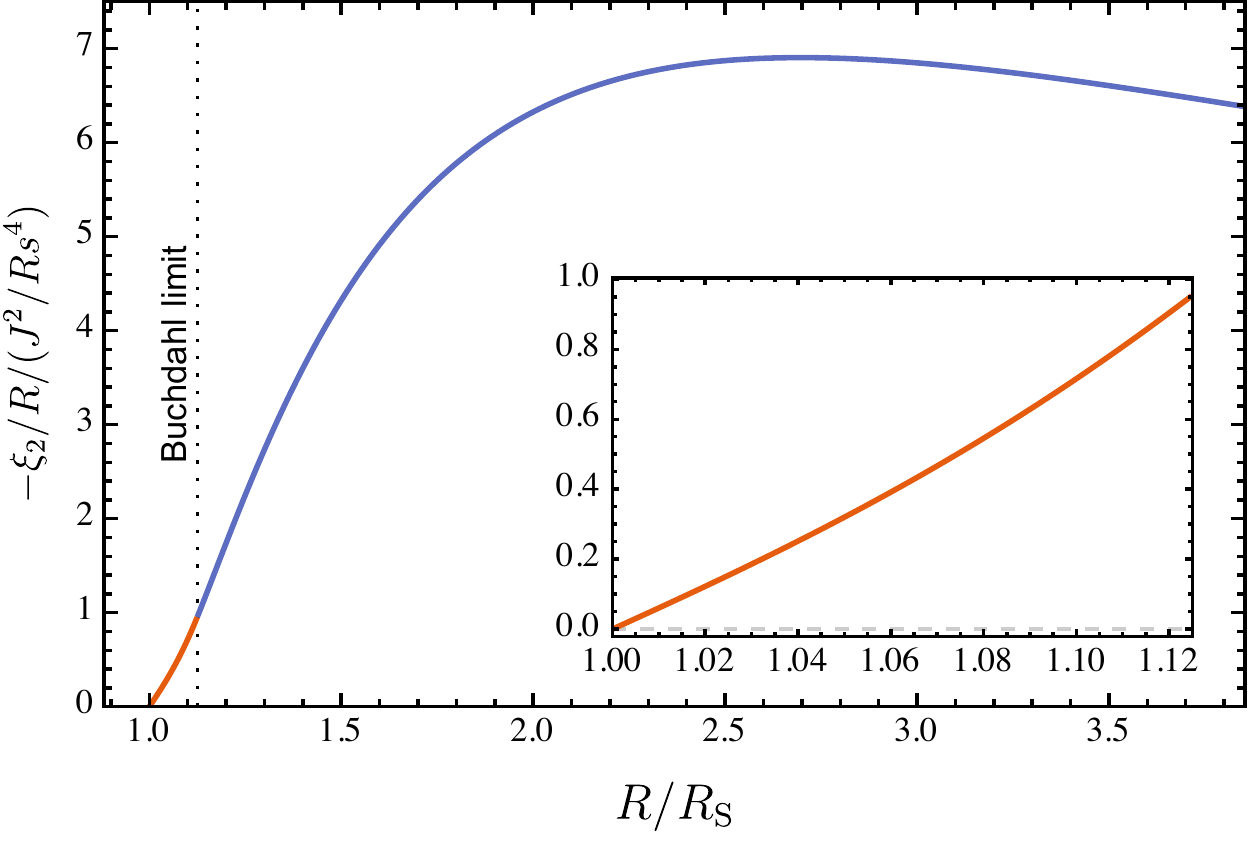}
\caption{The function $-\xi_2(R)/R$, which determines the $l=2$ deformation of the boundary, as a function of the parameter $R/\Rs$. We measure $-\xi_2(R)/R$ in the unit $J^2/\Rs^4$.}
\label{fig:xi2}
\end{figure}

In figure~\ref{fig:xi2} we display the $l=2$ boundary deformation function $-\xi_2(R)/R$, as a function of $R/\Rs$. As a verification test, we computed values of $-\xi_2(R)/R$ for configurations above the Buchdahl limit $R/\Rs > (9/8)$ which are in very good agreement with the results reported by \cite{Chandra:1974}. However, at the Buchdahl radius we obtain $\xi_{2}(\Rb)=-1.073(J^2/\Rs^3)$ which corrects the value reported by \cite{Chandra:1974}. We observe that in the gravastar limit, $-\xi_2(R)/R\to 0.$

\begin{figure}[ht]
\centering
\includegraphics[width=.6\linewidth]{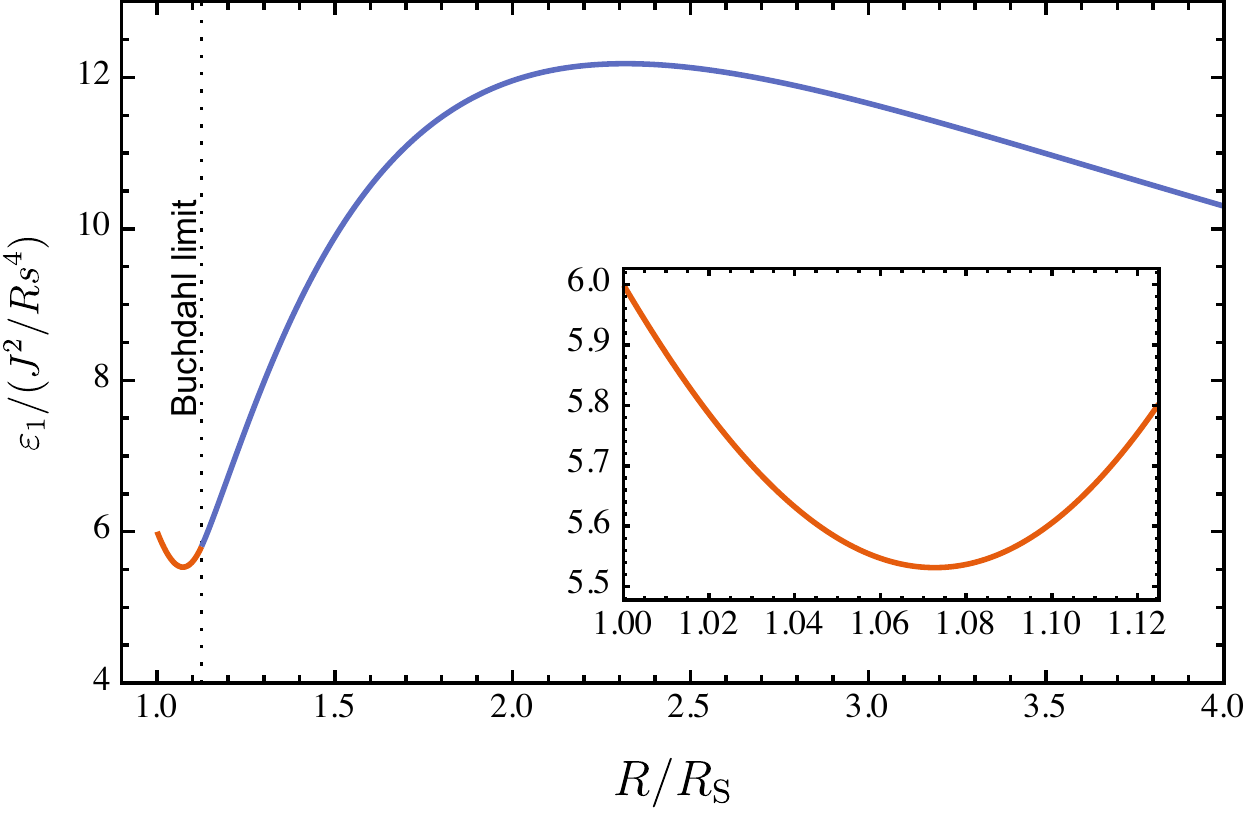}
\caption{Ellipicity of the isobaric surfaces $\varepsilon_{1}$, as a function of $R/\Rs$. In the gravastar limit $R\to\Rs$, $\varepsilon_{1}=6J^2/\Rs^4.$}
\label{fig:ellip}
\end{figure}

\begin{figure}[ht]
\centering
\includegraphics[width=.6\linewidth]{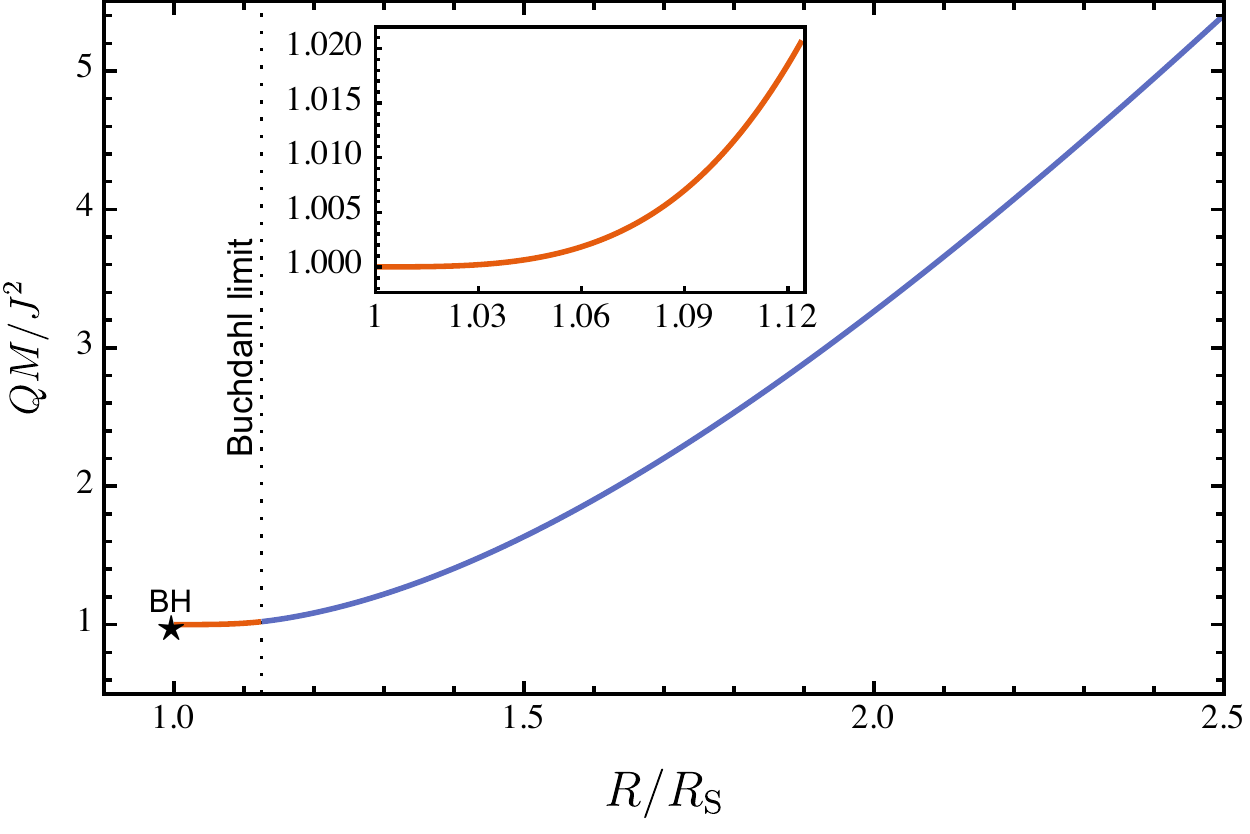}
\caption{Kerr factor $\widetilde{Q}=QM/J^2$, as a function of the parameter $R/\Rs$. The inset zooms into the regime beyond the Buchdahl limit. Observe the smooth and continuous approach of $\widetilde{Q}$ to the Kerr BH value, as $R\to\Rs$.}
\label{fig:Q}
\end{figure}

In figure~\ref{fig:ellip}, we present the ellipticity of the outer boundary $\varepsilon_1$, as given by equation \eref{ellipx}, as a function of the parameter $\ten$. We measure the ellipticity in units of $J^2/\Rs^4$. We observe that $\varepsilon$ shows a local maximum, $\approx~12.18$, at $\ten\approx 2.32$, in agreement with \cite{Chandra:1974}. There is also a local minimum, $\approx~5.53$, at $\ten\approx 1.08$ which is beyond the Buchdahl limit. At the Buchdahl radius we obtain $\varepsilon_{1}(\Rb)=5.8033$, which corrects the value reported by \cite{Chandra:1974}. In the gravastar limit $\ten\to 1$, we obtain
\begin{eqnarray}
\varepsilon=\frac{6J^2}{\Rs^4}.
\end{eqnarray}
Having determined all the perturbation functions for the slowly rotating Schwarzschild star in the gravastar limit, we proceed to determine the mass quadrupole moment $Q$, as given by \eref{Q}. In figure~\ref{fig:Q} we show the ‘Kerr factor' $\widetilde{Q}\equiv QM/J^2$, as a function of the parameter $R/\Rs$. This Kerr factor is relevant because it measures how much it deviates the exterior HT metric away from the Kerr metric ($\widetilde{Q}=1$). We observe how $\widetilde{Q}$ decreases monotonically as the compactness of the configuration increases. Our results for configurations above the Buchdahl limit, $\ten > (9/8)$, are in very good agreement with those in~\cite{Chandra:1974}. However, at the Buchdahl limit we obtain $\widetilde{Q}(\Rb)=1.0212$ which corrects the value reported by \cite{Chandra:1974}. In the limit as $R\to\Rs$, we observe that $\widetilde{Q}$ approaches quickly to 1. Thus, we obtain that in the gravastar limit
\begin{eqnarray}
Q=\frac{J^2}{M},
\end{eqnarray}
which agrees with the Kerr metric, within the approximation. Thus, we conclude that it is not possible to differentiate a slowly rotating Schwarzschild star, in the gravastar limit, from a Kerr BH, at the second-order in the angular velocity $\Omega$.
\begin{figure}[ht]
\centering
\includegraphics[width=.6\linewidth]{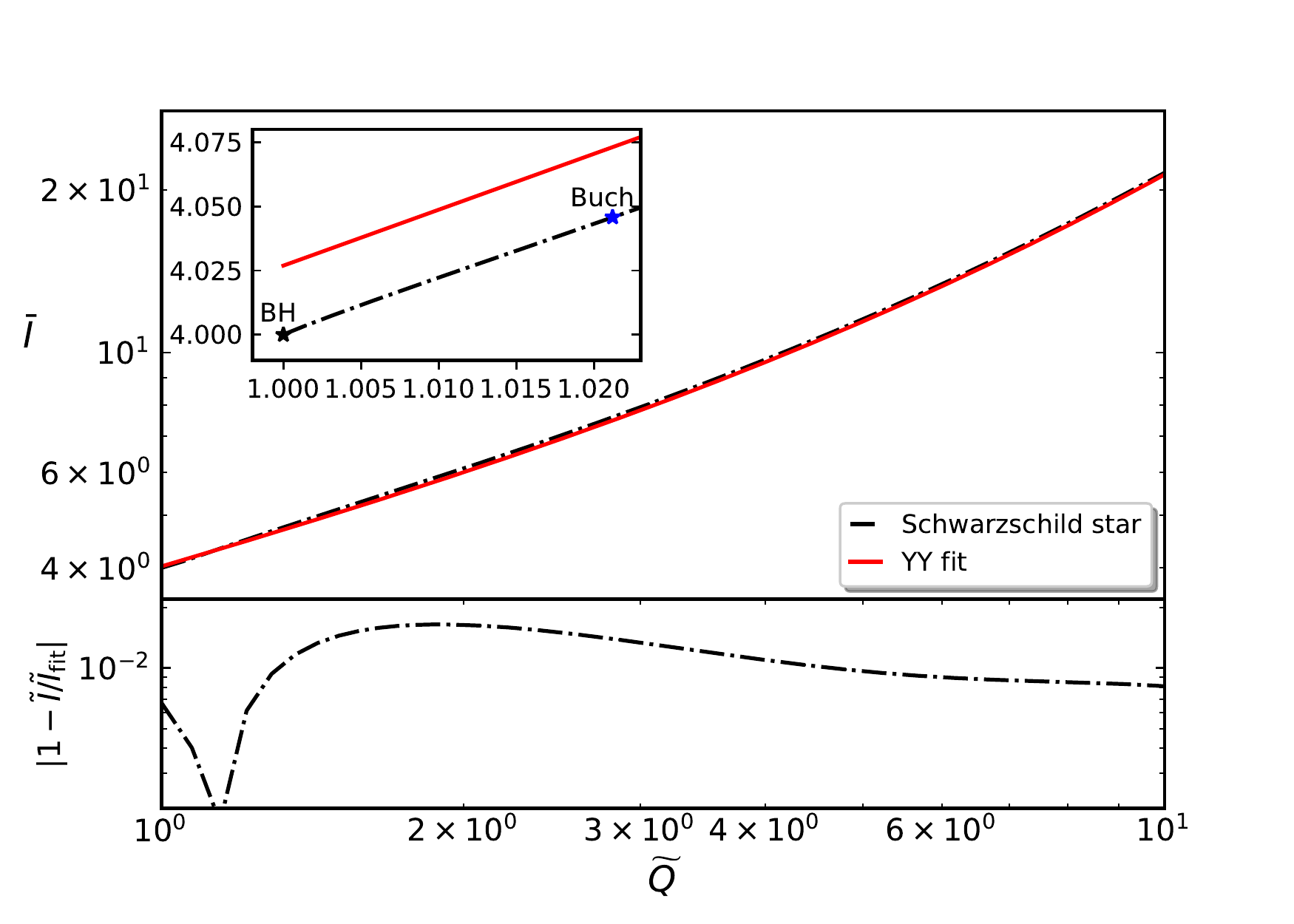}
\caption{$\bar{I}-\widetilde{Q}$ relation for the Schwarzschild stars below and beyond the Buchdahl limit (dashed black curve). We include the analytic fitting formula \eref{fitYY} proposed by Yagi and Yunes~\cite{Yagi:2016bktf} for realistic NSs (solid red curve). The inset zooms into the regime beyond the Buchdahl limit. The blue star indicates the Buchdahl bound, and the black star indicates the BH limit. In the bottom panel we show the fractional difference between the numerical results and the universal fit.}
\label{fig:IQ}
\end{figure}
Finally, from the results presented in figures \ref{fig:IN} and \ref{fig:Q}, in figure~\ref{fig:IQ} we display the $\bar{I}-\widetilde{Q}$ relation for slowly rotating Schwarzschild stars (dashed black curve). In the same figure, we also plot the universal fit of the $\bar{I}-\widetilde{Q}$ relations for realistic neutron stars (NSs) proposed by Yagi and Yunes (YY)~\cite{Yagi:2016bktf} which is given as follows
\begin{eqnarray}\label{fitYY}
\ln\bar{I} = a + b\ln(\widetilde{Q}) + c(\ln\widetilde{Q})^2 + d(\ln\widetilde{Q})^3 + e(\ln\widetilde{Q})^4,
\end{eqnarray}
where the fitting coefficients are: $a=1.393$, $b=0.5471$, $c=0.03028$, $d=0.01926$ and $e=4.434\times 10^{-4}$. 
In the inset we enhance the region beyond the Buchdahl limit. Note that the $\bar{I}-\widetilde{Q}$ relation for Schwarzschild stars approach exactly to the BH limit as $R\to\Rs$. We observe that deviations of the $\bar{I}-\widetilde{Q}$ relation for the Schwarzschild stars, with respect to the universal YY fit, are $\mathcal{O}(1)\%$. While \cite{Yagi:2013awa} considered the polytrope $n=0$ (corresponding to an incompressible fluid with  constant density) this shows that the fit is still reasonably accurate for Schwarzschild stars beyond the Buchdahl limit, despite the fact that the equations of state used to derive the fit in the first place are for lower density objects. 

\subsubsection{Inward extension of the quadrupole functions}

For the inward extension, $\varpi=0$ everywhere inside the infinite pressure surface $x_0$ is uniquely well behaved, as such the  behavior of the functions $h_2$ and $v_2$, inside $x_0$, becomes entirely dictated by the homogeneous equations \eref{v2h}, \eref{h2h}, while the behavior near $x_0$ is still specified by \eref{v2h_bc}, \eref{h2h_bc}. It is important to realize that for the homogeneous solution, $h_2^{(h)}\rightarrow0$ on $x_0$. Since in the region $0<x<x_0$ we only have the homogeneous solution, we are forced to accept
\begin{eqnarray}
    h_2(x\rightarrow-\kappa^-)=0.
\end{eqnarray}
 In practice we do not have $h_2(x\rightarrow-\kappa^+)=0$ for general cases (see left-hand panel of figure~\ref{fig:h2}), which implies $h_2$ may be discontinuous at the infinite pressure surface. Once again, a finite discontinuity of an $h$ perturbation function on $x_0$ is not a concern for $g_{tt}$ overall because $e^{2\nu}\rightarrow0$ there. However, we must have a continuous $k_2=v_2-h_2$ for continuous $g_{\theta\theta}$. The structure of \eref{v2h_bc} and \eref{h2h_bc}, together with the continuity of $k_2$ gives for the interior solution
 \begin{eqnarray}
C=\lim_{x\rightarrow-\ka^+}(k_2).\label{inQbound}
 \end{eqnarray}
Since $k_2$ typically does not go to zero as $x_0$ is approached from the outside (see left-hand panel of figure~\ref{fig:k2}) the quadrupole functions must have a nonzero behavior in the region $0<x<x_0$, unlike the monopole functions.

\begin{figure}[ht]
\includegraphics[width=.6\linewidth]{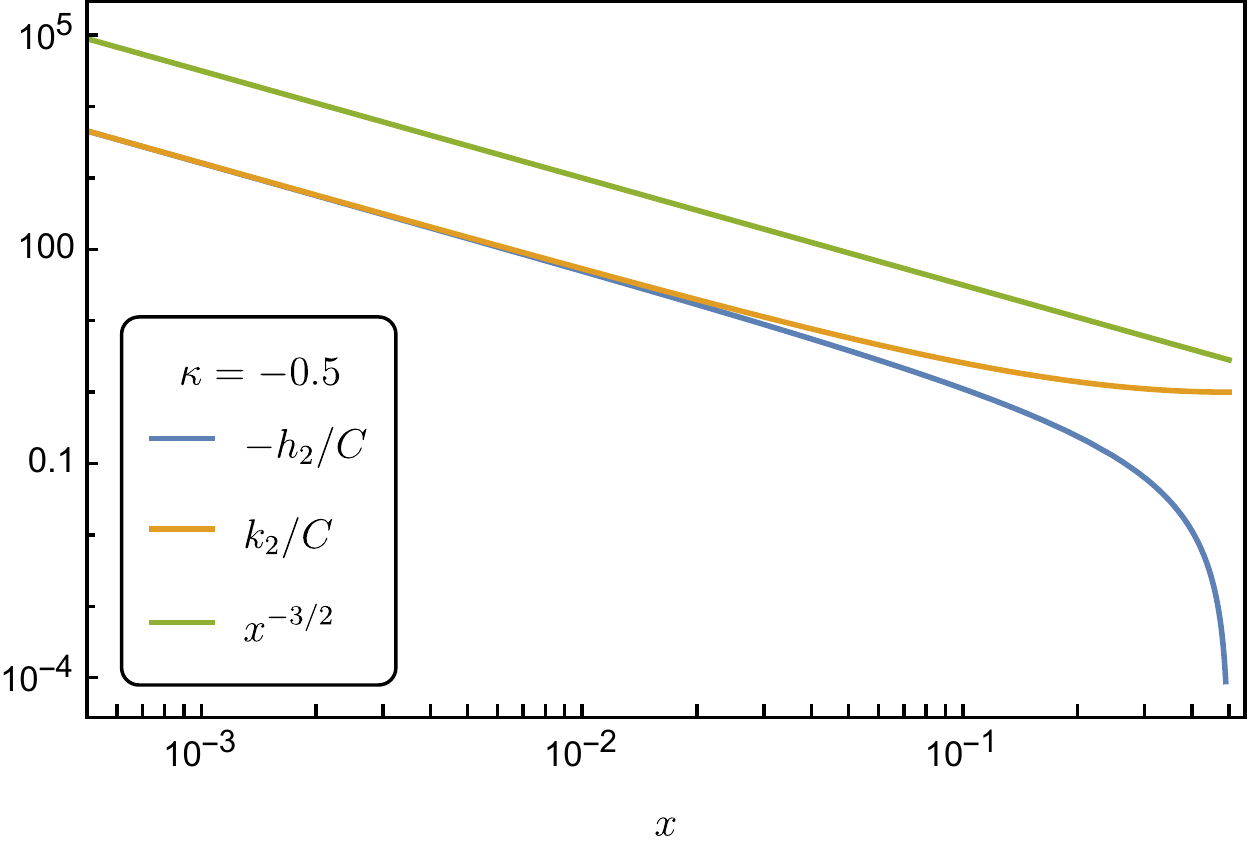}
\caption{ Log-Log plot of inward integration for $k_2$ and $h_2$ with asymptotic behavior highlighted for the example case of $\kappa=-0.5$. The function $k_2$ must be continuous across the infinite pressure surface and is generally nonzero, its value becoming the constant $C$ for the homogeneous solution. The function $h_2$ goes to zero as we approach the infinite pressure surface from the inside. Both functions have $x^{-3/2}$ divergences at the origin. }
\label{fig:qin}
\end{figure}

Unfortunately, inward integration of the quadrupole functions, given the boundary condition\eref{inQbound}, typically leads to divergences at the origin. It is especially noteworthy that the behavior, as we approach the origin numerically (see figure~\ref{fig:qin}), is 
\begin{eqnarray}
h_2, k_2\propto x^{-3/2}
\end{eqnarray}
which is equivalent to the $r^{-3}$ behavior at the origin for the analytic $k_2$ and $h_2$ terms derived in \cite{Beltracchi:2021lez}.


\section{Conclusions}
\label{Sect:6}

In this paper we have reconsidered the problem of a slowly rotating homogeneous star, or Schwarzschild star, when its compactness goes beyond the Buchdahl limit and approaches the gravastar limit $R\to\Rs$. For that purpose, we have employed the Hartle-Thorne (HT) framework for slowly rotating relativistic masses, at second order in the angular velocity $\Omega$. In our approach, we contracted adiabatically, in a quasi-stationary process, a Schwarzschild star by taking its compactness beyond the Buchdahl limit. By doing this process, we determined surface and integral properties associated to the region with positive pressure $x_0<x<x_1$. In the gravastar limit as $R\to\Rs$, $x_0\to x_1$, thus this positive pressure region becomes essentially a boundary layer where the pressure is \emph{positive}. Therefore, all the rotational properties of the configuration are associated with this boundary layer, while the interior de Sitter with negative pressure does not carry any angular momentum.

By imposing boundary conditions at the infinite pressure surface $x_0$, rather than the origin, we were able to integrate the perturbation functions in the the corresponding region $x_0<x<x_1$. The most important result of our study is that, in the gravastar limit, the perturbations and consequently quantities like the moment of inertia, and mass quadrupole moment, approach the corresponding Kerr metric values. Thus, within this approximation, it is not possible to differentiate a slowly rotating gravastar from a Kerr BH, by any measurements or observations in the exterior spacetime. Although we reached the same conclusions in \cite{Posada:2016xxx}, this study improves the one there where the study of the boundary conditions at $x_0$ was overlooked. However, in the relevant limit as $R\to\Rs$, the results are in agreement. 

 Our approach is different from the one presented by \cite{Beltracchi:2021lez}, where the authors considered a gravastar fully formed, and they computed the perturbations in the interior de Sitter with negative pressure. Considering that the HT equations, as formulated here, cannot be applied to the interior $p=-\epsilon$, the authors derived the corresponding structure equations \emph{ab initio}. We found that in the relevant limit when $R\to\Rs$ our results are in agreement with those in \cite{Beltracchi:2021lez}. Extension of the perturbation functions into the negative pressure region $0<x<x_0$, seems to lead to, in a best case scenario, divergences in the quadrupole perturbation functions at the origin. Interestingly, these divergences have qualitatively the same behavior $x^{-3/2}\propto r^{-3}$ as the corresponding quadrupole divergences at the origin found for the slowly rotating gravastar considered by \cite{Beltracchi:2021lez}.

 One avenue for future investigation, would be to consider other sub-Buchdahl systems in slow rotation and see how they compare to our model. In particular, while perfect fluid spheres with positive pressure are limited by the Buchdahl bound, spheres with anisotropic stress can be more compact \cite{Bowers:1974tgi, Lemaitre:1997gd, Ovalle:2019lbs}. Although the original HT approximation deals with perfect fluids, extensions of the HT method which allow for treatment of anisotropic stress have recently been pioneered by \cite{Pattersons:2021lci,Beltracchi:2022vvn}, therefore such a calculation may be possible.

 \section{Acknowledgements}
P.B. would like to acknowledge Paolo Gondolo for partnership in work regarding series solutions for $\varpi$ and divergences in $m_0$ in an unpublished project predating this one \cite{Beltracchi:2021}. We would also like to thank Cecilia Chirenti, Paolo Gondolo and Emil Mottola for valuable discussions. C.P. acknowledges the support of the Institute of Physics and its Research Centre for Theoretical Physics and Astrophysics, at the Silesian University in Opava.   

\appendix
\section{Series solution to the dragging equation}
\label{appendix1}

 Here we summarize the series solutions from \eref{drag} discussed by~\cite{Beltracchi:2021}. For that purpose, we introduce the new variable $z\equiv\kappa+x$ such that \eref{drag} takes the general form
\begin{eqnarray}
A(z)\varpi(z)'' + B(z)\varpi(z)' + C(z)\varpi(z)=0, \label{omegabardeqA}
\end{eqnarray}  
\noindent where 
\begin{eqnarray}
&A(z) = z(z-\kappa)(2+\ka-z),\\
&B(z) = \ka(2+\ka) + 3(1+\ka)z - 4z^2,\\
&C(z) = -4(1+\ka).
\end{eqnarray}
Let us investigate the regular singular point $z=0$. We can find a solution in terms of a Frobenius series as follows
\begin{eqnarray}
\varpi_1(z)=z^{2}\sum_{n=0}^{\infty}a_n z^n, \label{frob}
\end{eqnarray}
\noindent where the first few coefficients read
\begin{eqnarray}
a_{0}=1,\quad a_{1}=\frac{2(1+\ka)}{\ka(2+\ka)},\quad a_{2}=\frac{17+24\ka + 12\ka^2}{4\ka^2 (2+\ka)^2}.\label{omegabarcoefficientsA}
\end{eqnarray}
\noindent The coefficients obey a recurrence relation of the form
\begin{eqnarray}
i(i+2)\kappa(2+\kappa)a_{i} - (2i^2+5i-1)(1+\kappa) a_{i-1} + i(i+3)a_{i-2} = 0.
\end{eqnarray}
A second solution of $\varpi$ can be obtained in the form
\begin{eqnarray}
\varpi_{2}(z)=\sum_{n=0}^{\infty}b_n z^n - \frac{(1+\ka)^2}{[\ka(2+\ka)]^2}\log(z^2)\varpi_1(z), \label{omegabarseries2}
\end{eqnarray}
\noindent where the first few $b$ coefficients are
\begin{eqnarray}
&b_0=1,\quad b_1=\frac{4(1+\ka)}{\ka(2+\ka)},\quad b_2=0,~\nonumber\\
&b_3=-\frac{2 \left(9 \ka^3+27 \ka^2+19 \ka+1\right)}{3 \ka^3 (\ka+2)^3}.
\end{eqnarray}
The recurrence relation for these coefficients is
\begin{eqnarray}
    &-\frac{2 (4 i+13) (k+1)^3 a_{i+1}}{k^2 (k+2)^2}+\frac{2 (2 i+7) (k+1)^2 a_i}{k^2 (k+2)^2}+\nonumber\\
    &\frac{4 (i+3)
   (k+1)^2 a_{i+2}}{k (k+2)}+(i (2 i+13)+17) (k+1) b_{i+3}\nonumber\\
   &=(i+2) (i+5) b_{i+2}+(i+2) (i+4) k (k+2) b_{i+4}.
\end{eqnarray}
The most general possible solution to ~(\ref{omegabardeqA}) could then be written as
\begin{eqnarray}
\varpi_G=C_1 \varpi_1+C_2 \varpi_2 \label{omegabarG}.
\end{eqnarray}
Suppose we look at $\varpi_G$ at the infinite pressure surface. The $\varpi_1$ branch \eref{frob} goes to 0 as $z^2$. Although the other branch \eref{omegabarseries2} contains a term with a singular $\log$ factor, it is multiplied by the solution from the $a$ branch which goes to 0 fast enough to eliminate that term as $z\rightarrow0$, such that the highest order behavior is $\varpi_2\rightarrow b_0=1$; therefore
 \begin{eqnarray}
\lim_{z\rightarrow0}\varpi_G=C_2.
 \end{eqnarray}
 From \eref{m0inx} for $m_0$ we obtain
 \begin{eqnarray}
\frac{dm_0}{dx}=\frac{8 C_2^2 (-\ka)^{3/2} (\ka+1)^2 (\ka+2)^{3/2}}{3 (\ka+x)^3} + \mathcal{O}\left((x+\ka)^{-2}\right),\label{m0diverge}
 \end{eqnarray}
 \noindent which after integration will lead to a divergent $(x+k)^{-2}$ term \cite{Beltracchi:2021} that can only be cancelled by setting $C_2=0$. We can see from ~\ref{m0_bc} that $m_0$ remains finite at $x=-\kappa$ if we use only the $\varpi_1$ branch. Therefore, unless it is explicitly stated otherwise, we use 
 \begin{eqnarray}
     \varpi=\varpi_1.
 \end{eqnarray}

\section{Consistency check}
\label{appendix2}

As mentioned previously in the paper, we have additional codes which compute quantities both in the $(x,\ka)$ coordinates and in the $(r,R)$ coordinates. Here we examine the relative error defined by 
\begin{eqnarray}\label{ereq}
e(Z)\equiv\left\vert\frac{Z_{x\kappa}-Z_{rR}}{Z_{x\kappa}}\right\vert\,,
\end{eqnarray}
where Z is some quantity computed in the indicated coordinates. We compute the error of various quantities for 600 evenly spaced points between $R/\Rs=1$ and $R/\Rs=4$.

For the first order frame dragging effects we examine $\widetilde{\Omega}$ and $\varpi_1$. The average error in $\widetilde{\Omega}$ was $1.075\times 10^{-7}$, the maximum was $2.19\times 10^{-7}$ at $R/\Rs=3.7625$. The average error in $\varpi_1$ was $4.75\times 10^{-8}$, the maximum was $2.82\times 10^{-6}$ at the lowest point tested $R/\Rs=1.0025$ when $\varpi_1\rightarrow0$, which inflates the relative error as defined by \eref{ereq}.

For the monopole effects, we look at $\xi_0/J^2$ and $\widetilde{\delta M}$. For $\widetilde{\delta M}$, the average error is $5.34\times 10^{-6}$ and the maximum error is $1.48\times 10^{-4}$ which occurs at $R/\Rs=1.0325$. For $\xi_0/J^2$, the average error is $5.13\times 10^{-5}$ and the maximum error is $7.3\times 10^{-3}$, at the lowest point tested $R/R_S=1.0025$ when $\xi_0\rightarrow0$.

Finally, we can look at $\widetilde{Q}$ and $\xi_2/J^2$ as examples for the quadrupole sector. For $\xi_2/J^2$, the average error is $4.57\times 10^{-5}$ and the maximum error is $2.76\times 10^{-4}$, at slightly above the Buchdahl limit $R/\Rs=1.1275$. For $\widetilde{Q}$, the average error is $6.87\times 10^{-5}$ and the maximum error is $4.24\times 10^{-4}$ at $R/\Rs=2.3075$.

\section{Surface gravity parameter}
\label{appendix3}

In this appendix we discuss the surface gravity parameter for slowly rotating Schwarzschild stars. One convenient notation of the stationary axisymmetric metric in coordinates $(t,r,\theta,\phi)$ is
\begin{eqnarray}
ds^2 = -e^{2\Phi} dt^2 + e^{2\psi} \left(d\phi - \omega dt\right)^2 + e^{2\alpha} dr^2 + e^{2\beta} d\theta^2,
\label{axisymstat}
\end{eqnarray}
where $\Phi$, $\psi$, $\alpha$, $\beta$, and $\omega$ are functions of $r$ and $\theta$ only. The functions $\Phi$, $\psi$, and $\omega$ can be isolated as scalar functions due to the existence of the time and axial Killing vectors 
\begin{eqnarray}
&K_{(t)}^\mu=(1,0,0,0),\\
&K_{(\phi)}^\mu=(0,0,0,1).
\end{eqnarray}
We define additional vectors $l$ and $N$, which in these coordinates are
\begin{eqnarray}
&l^\mu=(1,0,0,\omega),\\
&N_\mu=(-1,-e^{\alpha-\Phi},0,0),
\end{eqnarray}
such that
\begin{eqnarray}
l=K_{(t)}+\omega K_{(\phi)},\quad N\cdot N=0,\quad N\cdot l=-1.    
\end{eqnarray}
With these auxiliary vectors, we can define two physically relevant scalar quantities, being the surface gravity parameter
\begin{eqnarray}\label{sg}
S_G = N_\mu l_\nu(\nabla^\mu l^\nu)=\frac{1}{2}e^{-\alpha-\Phi}\frac{\partial}{\partial r}e^{2\Phi},
\end{eqnarray}
and the angular momentum density parameter
\begin{eqnarray}\label{}
\mathcal{J} = -N_\mu l_\nu(\nabla^\mu K_{(\phi)}^\nu)=-\frac{1}{2}e^{2\psi-\alpha-\Phi}\frac{\partial \omega}{\partial r}.
\end{eqnarray}
These scalars are related to the Komar mass and angular momentum which can be defined as surface integrals at a given radius, or as the sum of a surface integral at a smaller radius and a volume integral of components of the energy-momentum tensor between the smaller and given radii \cite{Komar:1959,Beltracchi:2021zkt}
\begin{eqnarray}
M_K(r) & = \frac{1}{4\pi G}\int_{\partial V_+}\!(S_G + \omega \mathcal{J})\,dA \\
& = \int_V \sqrt{-g}\  \Big(\! \!-T^t_{\ t} + T^r_{\ r}  + T^\theta_{\ \theta}  + T^\phi_{\ \phi}  \Big)\, dr\,d\theta\,d\phi\, \\
& +\, \frac{1}{4\pi G} \int_{\partial V_-} \! (S_G + \omega \mathcal{J})\, dA,
\end{eqnarray}
\begin{eqnarray}
J_K(r) & = \frac{1}{8 \pi} \int_{\partial V_+}\! \! \mathcal{J} \, dA \\
& = \int_V  \sqrt{-g}\ T^t_{\ \phi} \,dr\,d\theta\,d\phi 
+\frac{1}{8 \pi } \int_{\partial V_-}\!  \!\mathcal{J} \, dA. 
\end{eqnarray}
For his perturbative framework, Hartle~\cite{Hartle:1967he} expanded the line element\eref{axisymstat} in the form \\
\eref{axialmetric}, with the following identifications
\begin{eqnarray}
&e^\Phi = e^{\nu(r)}\left[1 + h_0(r) + h_2(r) \,P_2(\cos \theta)\right],\\
&e^\psi = r \sin\theta\,\left[1 + k_2(r)\, P_2(\cos \theta)\right],\\
&e^\alpha = \left[1 + \frac{m_0 + m_2 P_2(\cos\theta)}{r- 2 m}\right]\left(\frac{r}{r-2m}\right)^{-1/2},\\
&e^\beta = r \left[1  + k_2(r) \, P_2(\cos \theta)\right],\\
&\omega=\omega(r),
\end{eqnarray}
where $\nu(r)$ corresponds to the nonrotating $g_{tt}$ metric element. When expanded to second order, the parameters $S_G$ and $\mathcal{J}$ give
\begin{eqnarray}
S_G =& e^{\nu}\Biggl[\nu'\left(1+h_0+h_2 P_{2}
- \frac{m_{0} + m_{2} P_{2}}{r-2m(r)}\right)\\
&+ h_0' + P_{2}(\cos\theta) h_2'\Biggr]{\left(\frac{r}{r-2 m(r)}\right)^{-1/2}},
\end{eqnarray}
\begin{eqnarray}
\mathcal{J} = -\frac{r^2 \sin ^2\theta e^{-\nu}}{2 \sqrt{\frac{r}{r-2 m(r)}}}\omega'.
\end{eqnarray}
\subsection{Surface gravity and angular momentum of Schwarzschild stars}
For the nonrotating Schwarzschild star, $\mathcal{J}=0$ and 
\begin{eqnarray}
S_G(\mathrm{in}) = -\frac{Mr}{R^3}\,\mathrm{sgn}\left(\sqrt{1-\frac{2 M r^2}{R^3}}-3 \sqrt{1-\frac{2 M}{R}}\right),
\end{eqnarray}
\begin{eqnarray}
S_G(\mathrm{out}) = \frac{M}{r^2},
\end{eqnarray}
where $S_G(\mathrm{in})$ changes sign at the infinite pressure surface, due to the sign function. However, it is somewhat convenient to write everything in the $x,\kappa$ variables such that
\begin{eqnarray}
S_G(\mathrm{in})=\mathrm{sgn}(\ka+x)\frac{\sqrt{x(2-x)}}{2\alpha}.
\end{eqnarray}
For the slowly rotating Schwarzschild star, the functions $S_G$ and $\mathcal{J}$ take the form
\begin{eqnarray}\label{sgx}
S_G=&\frac{\sqrt{x(2-x)}}{2\alpha}~\Bigg[\mathrm{sgn}(\ka+x) \Bigg(1+h_0+h_2 P_{2}\left(\cos\theta\right)\nonumber\\
&- \frac{m_0 + m_2 P_{2}\left(\cos\theta\right)}{\alpha(1-x)^2 \sqrt{x(2-x)}}\Bigg)+ \vert x+\kappa\vert\left(h_0' + P_{2}(\cos\theta) h_2'\right)\Bigg],
\end{eqnarray}
\begin{eqnarray}\label{Jx}
   \mathcal{J}=-\frac{\alpha\left[x(2-x)\right]^{3/2} \sin^2\theta}{\vert \ka+x \vert}\omega',
\end{eqnarray}
where the prime now denotes derivative with respect to $x$. Inside the infinite pressure surface $x_0$, we have
\begin{eqnarray}
&\varpi = 0,\quad h_0=h_c(\mathrm{in}),\nonumber\\
&m_0 = 0,\quad m_2,h_2=O(x+\kappa)^2 \nonumber.
\end{eqnarray}
Thus, taking the limit of \eref{sgx} and \eref{Jx}, from below, i.e. $x\to -\ka^{-}$, we get
\begin{eqnarray}
&S_G(x \rightarrow -\kappa^-) = -\frac{\sqrt{-\kappa(2+\kappa)}}{2\alpha}\left[1+h_c(\mathrm{in})\right],\\
&\mathcal{J} = 0.
\end{eqnarray}
As we approach $x_0$ from above, the perturbation functions take the behaviors
\begin{eqnarray}
\varpi &= \varpi_0(x+\kappa)^2,\quad h_0=h_c(\mathrm{out}),\nonumber\\
h_2 &= -\frac{4}{3}\left[\ka(\ka+2)\right]^3 (\alpha\varpi_0)^2\nonumber.
\end{eqnarray}
The function $m_2$ can be found from $h_2$ and $\varpi$, but it is rather long so we do not write it down. Taking the limit from above, i.e. $x\to -k^{+}$, we obtain
\begin{eqnarray}
 & S_G(x\rightarrow-\ka^+) = \frac{\sqrt{-\ka(2+\ka)}}{2\alpha}\left[1+h_c(\mathrm{out})\right],\\
 & \mathcal{J}(x\rightarrow-\kappa^+) = 2\alpha\left[-\ka(\ka+2)\right]^{3/2} \varpi_0 \sin^2\theta.
\end{eqnarray}
Note that there is a discontinuity in $\mathcal{J}$ at the infinite pressure surface $x_0$, which implies that there is a $\delta$ function in $\sqrt{-g}~T^t_{~\phi}$, so there is angular momentum density carried on $x_0$. Likewise, there is also a $\delta$ function behavior in $\sqrt{-g}(T^\mu_{~\mu})$ associated with the discontinuity in $S_G+\omega\mathcal{J}$. The discontinuity in $\mathcal{J}$ is of the order $\Omega$ and the discontinuities in $S_G$ and $S_G+\omega\mathcal{J}$, generally consist of order $\Omega^0+\Omega^2$, although $S_G$ may not have any $\Omega^2$ discontinuity if $h_{c}(\mathrm{in})=-h_{c}(\mathrm{out})$. On the other hand, if $h_c(\mathrm{in})=h_c(\mathrm{out})$, then $S_G$ has equal and opposite magnitude on each side of the surface which is the case for the nonrotating Schwarzschild star.
\section*{References}
\bibliographystyle{iopart-num.bst}
\bibliography{Sstar_rot_revised}
\end{document}